\documentclass[twoside,11pt]{5ssmmp}

\usepackage{amsfonts}
\usepackage{amssymb}
\usepackage{amsmath}
\usepackage{fancyhdr}    
\usepackage{equations}
\usepackage{epic}
\usepackage{eepic}
\usepackage{epsf}
\usepackage{graphicx}
\usepackage{rotating}

\usepackage{bm}

\graphicspath{{./Figs/}}

\def\Black{}
\def\Blue {}
\newcommand\myMath[1]{${#1}$}

\begin{document}
 \baselineskip=11pt
 \title{Inevitability and Importance of Non-Perturbative
 Elements in Quantum Field Theory\hspace{.25mm}
\thanks{\,Work supported by the presidential grant
 Scientific School 3810.2010.2,
 RFBR grants No. 08-01-00686, 09-02-01149, and 11-01-00182,
 cooperative grant BRFBR--JINR (contract No.~F10D-001)
 and the Heisenberg--Landau Program (grant 2010--2011).}}
 \author{\textbf{Alexander\,P.\,Bakulev}\hspace{.25mm}
 \thanks{\,E-mail address: bakulev@theor.jinr.ru}
  \\ \normalsize{Bogoliubov Lab. Theor. Phys., JINR (Dubna,
 Russia)} \vspace{2mm}
  \\ \textbf{Dmitry\,V.\,Shirkov}\hspace{.25mm}
      \thanks{This author is responsible mainly for the text of Section 1.}
  \\ \normalsize{Bogoliubov Lab. Theor. Phys., JINR (Dubna, Russia)}}
 \date{\today}
 \maketitle

\begin{abstract}
 The subject of the first section-lecture is
 concerned with the strength and the weakness
 of the perturbation theory (PT) approach,
 that is expansion in powers of a small parameter $\alpha$,
 in Quantum Theory.
 We start with outlining a general troublesome feature
 of the main quantum theory instrument,
 the perturbation expansion method. The striking
 issue is that
 \textit{perturbation series in powers of
 $\alpha \ll 1$ is not a convergent series}. The formal reason
 is an essential singularity of quantum amplitude (matrix
 element) $C(\alpha)$ at the origin $\alpha=0\,$.
 In many
 physically important cases one needs some alternative means of
 theoretical analysis. In particular, this refers to perturbative
 Quantum Chromodynamics (pQCD) in the low-energy domain.\\
\indent In the second section-lecture,
 we discuss the approach of Analytic Perturbation Theory (APT).
 We start with a short historic preamble and
 then discuss how combining
 the Dispersion Relation
 with the Renormalization Group (RG) techniques
 yields the APT
 with \myMath{\displaystyle e^{-1/\alpha}} nonanalyticity.
 Next we consider the results of APT applications
 to low-energy QCD processes
 and
 show that in this approach the fourth-loop contributions,
 which appear to be on the asymptotic border
 in the pQCD approach,
 are of the order of a few per mil.
 Then we note that using the RG in QCD
 dictates the need to use the Fractional APT (FAPT)
 and describe its basic ingredients.
 As an example of the FAPT application in QCD
 we consider the pion form factor \myMath{F_\pi(Q^2)} calculation.
 At the end, we discuss the resummation of nonpower series
 in {(F)APT} with application to the estimation
 of the Higgs-boson-decay width \myMath{\Gamma_{H\to\overline{b}b}(m_H^2)}.
\end{abstract}

\section{Strength and weakness of perturbative QFT}
 \label{sec:Lect1}
  \subsection{Essential singularity at \myMath{\alpha=0} and functional integrals}
  In quantum mechanics and QFT we have a lot of successful
perturbative calculations.
Practically, Perturbation Theory is a synonym of Quantum Theory.
Feynman diagrams became a symbol of QFT.
Nevertheless, perturbative power expansion of the quantum amplitude
\myMath{C(\alpha)}
is not convergent.
 \begin{center}
 \framebox{{Feynman Series \myMath{\sum c_k\alpha^k}
   is not Convergent !}}
 \end{center}
The reasons for this behavior are rather simple.
First, remind the Dyson argument~\cite{Dyson52}.
In QED the substitution \myMath{\alpha \to -\alpha}
is equivalent\footnote{
Here \myMath{\alpha={e^2}/(4\pi)} is the QED expansion parameter,
the fine structure constant.}
to \myMath{e\to i\,e}
and due to \vspace{-2mm}
\begin{eqnarray}
 \label{eq:S.T-exp}
  S = T(e^{i\!\int L_\text{int}(x)\,dx})
    = T(e^{i\,e\!\int j_\mu A^\mu\,dx})
\end{eqnarray}
this change destroys hermiticity of Lagrangian
 and unitarity of S-matrix.
Hence, in the complex \myMath{\alpha} plane,
the origin cannot be a regular point,
instead, one has an essential singularity at \myMath{\alpha=0}.

Second, in QFT one meets factorial growth of coefficients \myMath{c_k \sim k!}
and this is
due to an ill-posed problem.
A small parameter \myMath{g} standing at the highest nonlinearity
is an indispensable attribute of quantum perturbation.
Indeed, we quantize a linear system as a set of oscillators.
Only after that we account nonlinear terms
\myMath{\sim g \ll 1} as small perturbations.
However, nonlinearity usually changes the equation seriously ---
new solutions appear.

The most general way to analyze the issue is
to use
the functional integral representation.
Here, for illustration purposes only,
we consider the analog
(often called ``0-dimensional'')
of the scalar field theory \myMath{g \varphi^4}
\begin{subequations}
\begin{eqnarray}
 \label{eq:int.0D}
  I(g)=\int\limits_{\!\!-\infty}^{\,\,\infty} e^{-x^2-gx^4}\,d x\,.
\end{eqnarray}
Expanding it in power series
\begin{eqnarray}
 \label{eq:I-series}
   I(g) &=& \sum_{k=0}(-g)^k I_k\quad \text{with} \quad
   I_k  = \frac{\Gamma(2k+1/2)}{\Gamma(k+1)}\bigg|_{k\gg1}\,\,\to 2^k\,k!\,,
\end{eqnarray}
one arrives~\cite{KaSh80} at factorially growing coefficients.
Meanwhile, \myMath{I(g)}
can be expressed via special MacDonald function
\begin{eqnarray}
 \label{eq:I-McDon}
 I(g) =
  \frac{1}{2\,\sqrt{g}}\,e^{1/8g}\,
 K_{1/4}\left(\frac{1}{8g}\right)
\end{eqnarray}
with known analytic properties in the complex $g$ plane:
It is a four-sheeted function  analytical
in the whole complex plane
besides the cut
from the origin \myMath{g=0} along the whole negative semiaxes.
At the origin, it has an essential singularity
$\displaystyle{e^{-1/8g}}$
and in its vicinity on the first Riemann sheet
it can be written down
in the Cauchy integral form:
\begin{eqnarray}
 \label{eq:I-CInt}
  I(g) =
   \sqrt{\pi}
   - \frac{g}{\sqrt{2\,\pi}}\!
      \int_0^{\infty}
       \frac{d\gamma\,e^{-1/4\gamma}}{\gamma(g+\gamma)}\,.
\end{eqnarray}
\end{subequations}
As far as the origin is not an analytical point,
the power Taylor series (\ref{eq:I-series})
has no convergence domain
for real positive \myMath{g} values.
This is in concert with factorial growth
of power expansion coefficients.
The power series is not valid also for negative $g$ values
--- in accordance
 with Dyson's reasoning.

 Besides, via 0-dimensional analog of functional integral
(\ref{eq:int.0D})
one can illustrate the analysis
of the $I(g)$ analytic properties in the complex $g$ plane
by the steepest-descent method
which was devised\footnote{See, e.g., Section 2 in the paper\cite{KaSh80} and references therein.}
for the functional integral representation.
Then it is  possible to prove~\cite{Lip76}
factorial growth
of expansion coefficients
in the \myMath{\phi^4} scalar and a few other QFT models.
These results have been anticipated
in 1952--53~\cite{HuThPet52}
just after Dyson's paper.

The same singularity structure
${\displaystyle\sim \exp(-1/g)}$ was established
by improving the perturbative result
using two other nonperturbative methods:
\myMath{Q^2}-analyticity (from Dispersion Relations),
and Renormalization Invariance~\cite{ShiDKS77}:
\vspace{-3mm}
\begin{eqnarray}
 \label{eq:DR.RGI.Sing}
  f_\text{pert}(Q^2,g)
  = 1 - \beta_0\,g\,\ln(Q^2)
  \ \to \
  f_\text{imp}(Q^2\,e^{-1/\beta_0\,g})\,.
\end{eqnarray}

Henry Poincar\'e analysis of Asymptotic Series
(AS) properties at the end of the XIX century
can be summarized as follows:\vspace*{3mm}

\noindent
\begin{tabular}{lr}\!\!\!\!\!
 \begin{minipage}{0.48\textwidth}
  \includegraphics[width=\textwidth]{
   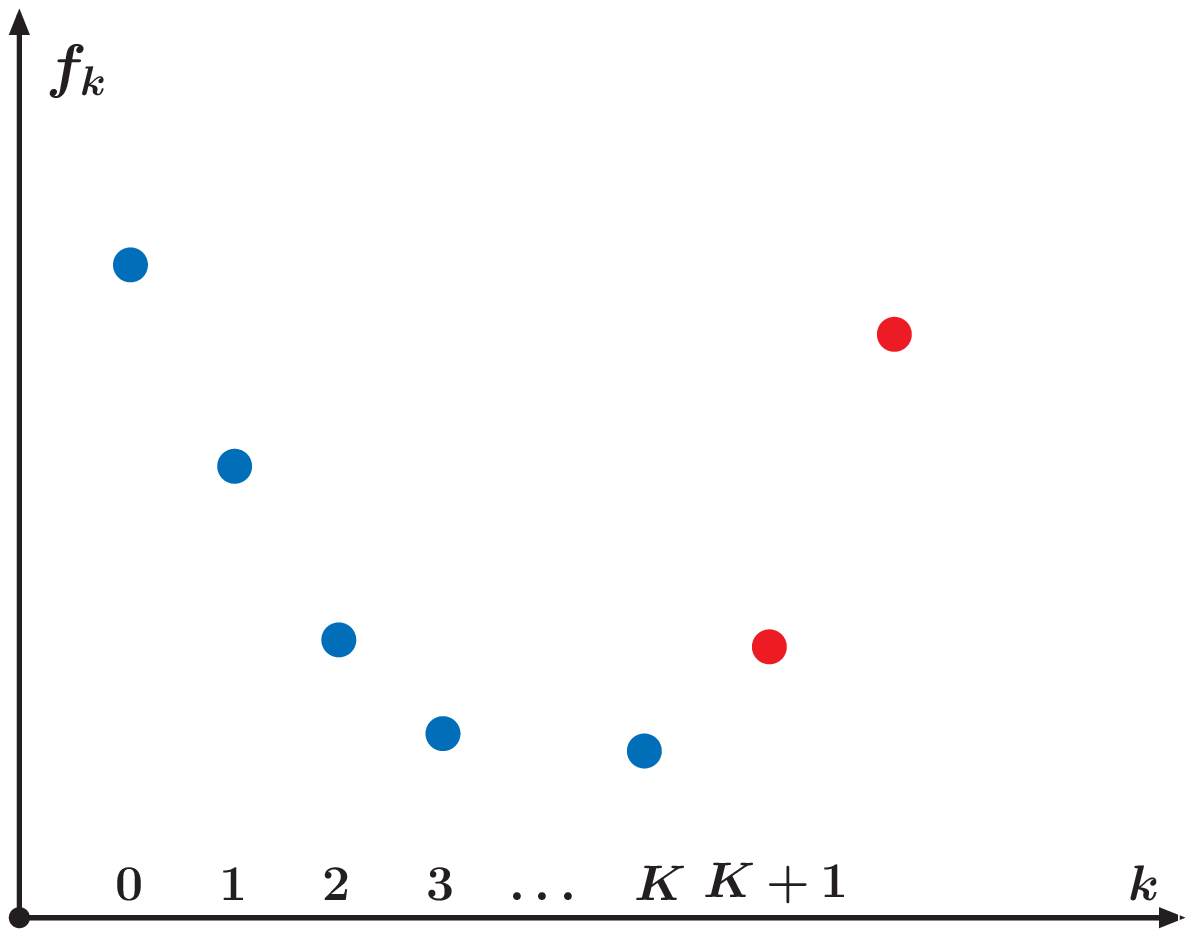}
  \end{minipage}
  &\!\!\begin{minipage}{0.475\textwidth}
 The truncated AS can be used for obtaining
  quantitative information on expanded function.
  Here, \underline{the error} of approximating $F(g)$ by
  first $K$ terms of expansion,
  {\small $ F(g)\to F_K(g)=\Sigma_{k\leq K}\,f_k(g)$}
  is equal to \underline{the} \underline{last detained term}
  $f_K(g)$. \\
  This yields to the existence of a \textit{lower limit of
 possible accuracy} for the given $g$ value (in contrast
 to convergent series!).
  \end{minipage}
  \end{tabular}\vspace*{2mm}

\indent To elucidate the phenomenon, take a power AS,
$f_k(g)=f_k\,g^k$ with factorial growth $f_k\sim k!$
of expansion terms $f_k(g)$
which cease to diminish at \myMath{k=K\sim 1/g}\,.
For $k\geq K+1$ truncation error starts to grow!

  This yields the natural {best possible accuracy}
of an AS at a given value of expansion parameter.
From the explicit illustration for the function $I(g)$
(\ref{eq:int.0D})
with AS (\ref{eq:I-series}),
presented in Table~\ref{tab:I(g)},
one can see that the optimum values of truncation number
$K=K_*\simeq1/(2g)$ (in dark blue)
provides us with the best possible accuracy
(dark blue in the last column).
Indeed, an attempt to account a couple of extra terms
(the second and the forth lines)
results in drastic rise of the error!

\begin{table}[!h]
  \centering
{\small
\begin{tabular}{|c||c||c|c||c|c|c|}\hline
\myMath{g\vphantom{^|_|}}
      &\myMath{K}
             &\myMath{(-g)^{K}\,I_{K}}
                    &\myMath{(-g)^{K+1}\,I_{K+1}}
                           &\myMath{I_{K}(g)}
                                  &\myMath{I(g)}
                                         &\myMath{\Delta_K I(g)}\\ \hline\hline
\myMath{0.07}
      &\myMath{\Blue\bm{7}\Black}
             &\myMath{-0.04 (2\%)}
                    &\myMath{+0.07 (4.4\%)\vphantom{^|_|}}
                           &\myMath{1.674}
                                  &\myMath{1.698}
                                         &\myMath{\Blue\bm{1.4\%}\Black} \\ \hline
\myMath{0.07}
      &\myMath{9}
             &\myMath{-0.17 (10\%)}
                    &\myMath{+0.42 (25\%)\vphantom{^|_|}}
                           &\myMath{1.582}
                                  &\myMath{1.698}
                                         &\myMath{7\%} \\ \hline
\myMath{0.15}
      &\myMath{\Blue\bm{2}\Black}
             &\myMath{+0.13 (8\%)}
                    &\myMath{-0.16 (10\%)\vphantom{^|_|}}
                           &\myMath{1.704}
                                  &\myMath{1.639}
                                         &\myMath{\Blue\bm{4\%}\Black} \\ \hline
\myMath{0.15}
      &\myMath{4}
             &\myMath{+0.30 (18\%)}
                    &\myMath{-0.72 (44\%)\vphantom{^|_|}}
                           &\myMath{1.838}
                                  &\myMath{1.639}
                                         &\myMath{12\%} \\ \hline
 \end{tabular}\vspace*{+1pt}}
 \caption{\small The last detained ($(-g)^{K}\,I_{K}$) and the first
  dismissed ($(-g)^{K+1}\,I_{K+1}$) contributions to the power series (\ref{eq:I-series})
  in comparison with the exact value (\ref{eq:I-McDon}) and the approximate result $I_K(g)$,
  with $K$ being the truncation number and $\Delta_K I(g)$ --- the error of approximation.
 \label{tab:I(g)}} \end{table}\vspace*{-2mm}
\noindent Thus, one has $\,K_*(g=0.07)=7\,$
and $\,K_*(g=0.15)=2\,$.
It is not possible at all to get the 1\% accuracy
for $g=0.15\,$.

  In QED this `divergence menace' is not actual,
as the real expansion parameter is quite small:
\myMath{\alpha/\pi\sim 1/(137\pi) \sim 2\cdot10^{-3}\,}.
At the same time, in perturbative QCD (pQCD)
the expansion parameter below 5--10 GeV is not very small:
\myMath{\alpha_s(Q)\sim 0.2-0.3}.
Then for an observable
$A_\text{QCD}=\sum\nolimits_k a_k(\alpha_s)^k$
with \ \myMath{a_k\sim k!}
and with critical order \ \myMath{K\sim 3-5}
the danger of the pQCD series explosion is actual,
see below in Table~\ref{tab:PT}.

 Hence, a practical Non-Perturbative approach is of utmost importance.
In Section~2, we concentrate on the so-called
Analytic Perturbation Theory (APT),
a closed scheme,
devised in the late 90s\cite{APTbasic}.
It combines information from PT
with two other nonperturbative methods ---
Analyticity and Renormalization Group (RG).
Having this in mind,
we outline now the RG approach.
Then ideas of the Dispersion Relation method
will be presented.
\subsection{RG transformation}
 Consider transformation
\myMath{R_t\left[\mu_i\to\mu_k\,,\,g_i\to g_k\right]}
as \textit{operation} with continuous positive
parameter \myMath{t},
acting on a \textit{group element}
\myMath{\mathcal G_i = \mathcal G(\mu_i,g_i)},
specified  by 2 coordinates  \myMath{\mu_i}
and \myMath{g_i}\footnote{For a more detailed exposition
of this material kindly address to~\cite{RG-00}.}.
This operation
\begin{eqnarray}
 \label{eq:RG}
  R_t\cdot\mathcal G_i
   = \mathcal G_k\sim R_t
      \left\{\mu_i \to \mu_k=t\mu_i,\,
                 g_i \to g_k =\overline{g}(t,g_i)
      \right\}
\end{eqnarray}
 contains dilatation of \myMath{\mu} and
functional transformation of \myMath{g_\mu}\,.
The \myMath{R_t} group structure
\myMath{R_t\,R_\tau = R_{t\,\tau}}
is provided by the functional equation (FEq)
\begin{eqnarray}
 \label{eq:RG.FEq}
  \overline{g}(\tau\,t,g)
   = \overline{g}\left(\tau,\overline{g}(t,g)\right)\,.
\end{eqnarray}
Indeed, if one puts \myMath{x=\tau\,t},
then its LHS describes \myMath{R_{\tau t}}
acting on
\myMath{g},
\myMath{R_{\tau\,t}\,g =} \myMath{\overline{g}(\tau\,t,g)}\,,
while the RHS one corresponds to the two-step procedure:\\
\myMath{R_\tau\otimes R_t\,g= R_\tau\,\overline{g}(t,g)
   =\overline{g}\left(\tau,\overline{g}(t,g)\right)}\,.
We see that a combination of the two lines results in Eq.\,(\ref{eq:RG.FEq}),
providing the group composition law
\myMath{R_{\tau t}=R_\tau\otimes\,R_t}.
Thus, the operation \myMath{R_t} forms
a continuous
Sophus Lie (1880) group of transformations.

 The RG symmetry and RG transformation are close
to the notion of {self-similarity},
well-known in Mathematical Physics since the end of the XIX century.
The {Self-Similarity Transformation} (SST)
is a simultaneous power scaling of arguments
\myMath{z=\{x, t,\ldots\}}
and functions \myMath{V_i(x,t,\ldots)}
\begin{eqnarray}
 \label{eq:SST}
  S_\lambda:
  \left\{\,x \to \lambda\,x\,,\,t \to
  \lambda^a\,t\,
  \right\}\,;\quad
 \left\{V_i(z)\to V_i'(z^\prime)
     = \lambda^{\nu_i} V_i(z^\prime)
  \right\}
\end{eqnarray}
Below, we call it the \textit{Power Self-Similarity} (PSS) transformation.

The general solution of
  \myMath{\overline{g}(xt,g)=\overline{g}(x,\overline{g}(t,g))}
depends on an arbitrary one-argument function
see below Eq.\,(\ref{eq:RGFEq-sol}).
Here, we look for a partial solution,
linear in the second argument
\myMath{\overline{g}(x,g)= g\cdot f(x)}.
The function \myMath{f(x)} satisfies
simple FEq
\myMath{f(xt)=f(x)\cdot f(t)}
with general solution:
\myMath{f(x)=x^{\nu}} and \myMath{\overline{g}(x,t)=g\cdot x^\nu}.
Thus, RG transformation is reduced to the PSS one,
$$R_t\to\{x\to x\cdot t,~
           g\to g\cdot t^\nu\}=S_t.$$

The PSS transformation
\myMath{R_t\to S_t=\{x\to x\cdot t,~g\to g\cdot t^\nu\}}
is a special case of the RG one.
That is, in the RG case instead of the power law \myMath{t^\nu},
one has
\textit{arbitrary functional} dependence.
Hence, one can consider the RG transformation
as a \textit{functional} generalization of the PSS one.
It is natural then to treat them
as transformations
of functional scaling or
\textit{Functional Self-Similarity} (FSS) transformation.
In short \text{RG \myMath{\equiv} FSS}.

\begin{figure}[h!]
\centerline{\includegraphics[width=0.83\textwidth]{
 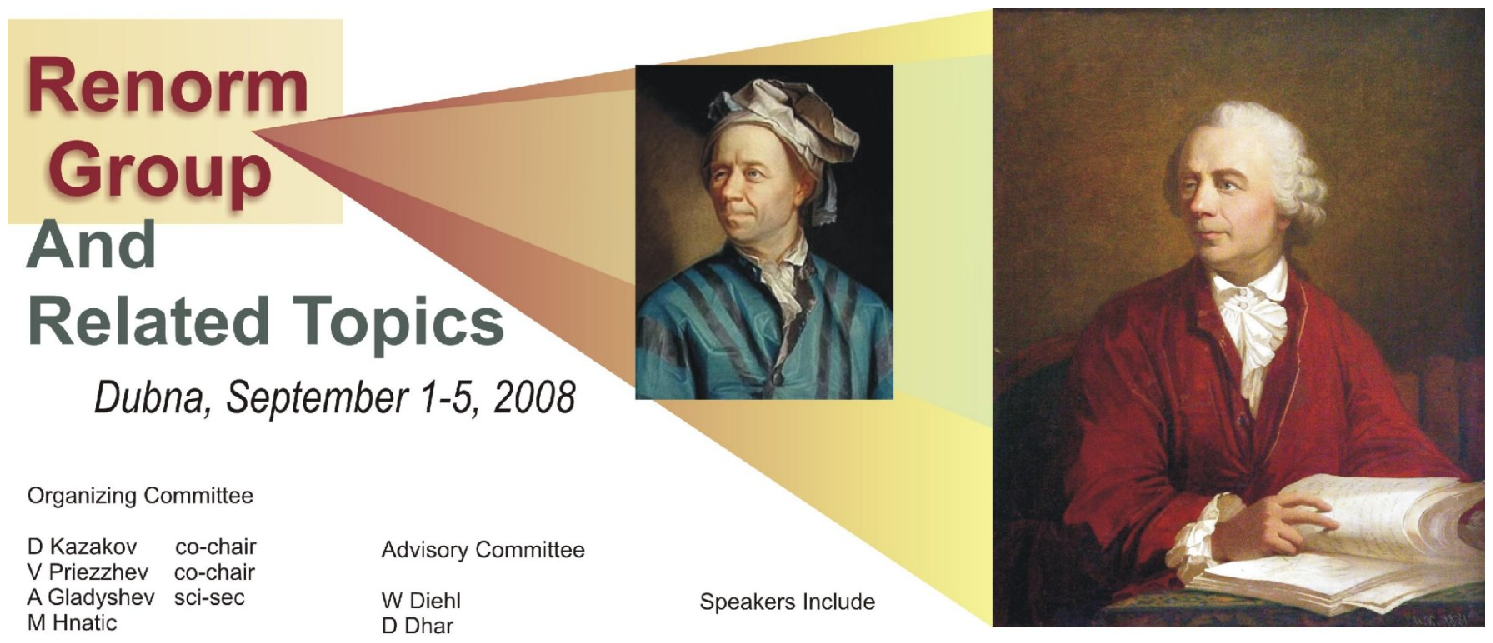}}
\caption{\small The project for poster of the 2008 RG
 Conference in Dubna. \label{fig:euler}}
\end{figure}

Here we illustrate our statements by historical analogies.
One realizes that
\textit{scaling}
= change of scale
= \textit{proportional change of sizes}.
This notion implies that \textit{Bogoliubov Renormalization Group}
(BRG)
= distorted scaling
= \textit{change of scale with continuous change of some details}.
This is illustrated by Fig.\,\ref{fig:euler}:
in both paintings one can recognize the same person,
Leonard Euler but details are different\footnote{
The original (smaller in Fig.\,\ref{fig:euler})
portrait was painted by E.~Handmann in 1756
and is stored at Basel University,
whereas the larger and more official one was painted
by I.~K\"{o}nig at the request
of the Russian Academy of Sciences
(due to its 150-year anniversary)
in 1875 as a copy from the original portrait.
This newer portrait was lost during Russian revolution
and discovered in a curiosity shop in 1972
by Georgy Sergeevich Golitsyn~\cite{Gol07}.}.

\begin{figure}[ht]
 \centerline{\includegraphics[height=0.34\textwidth]{
  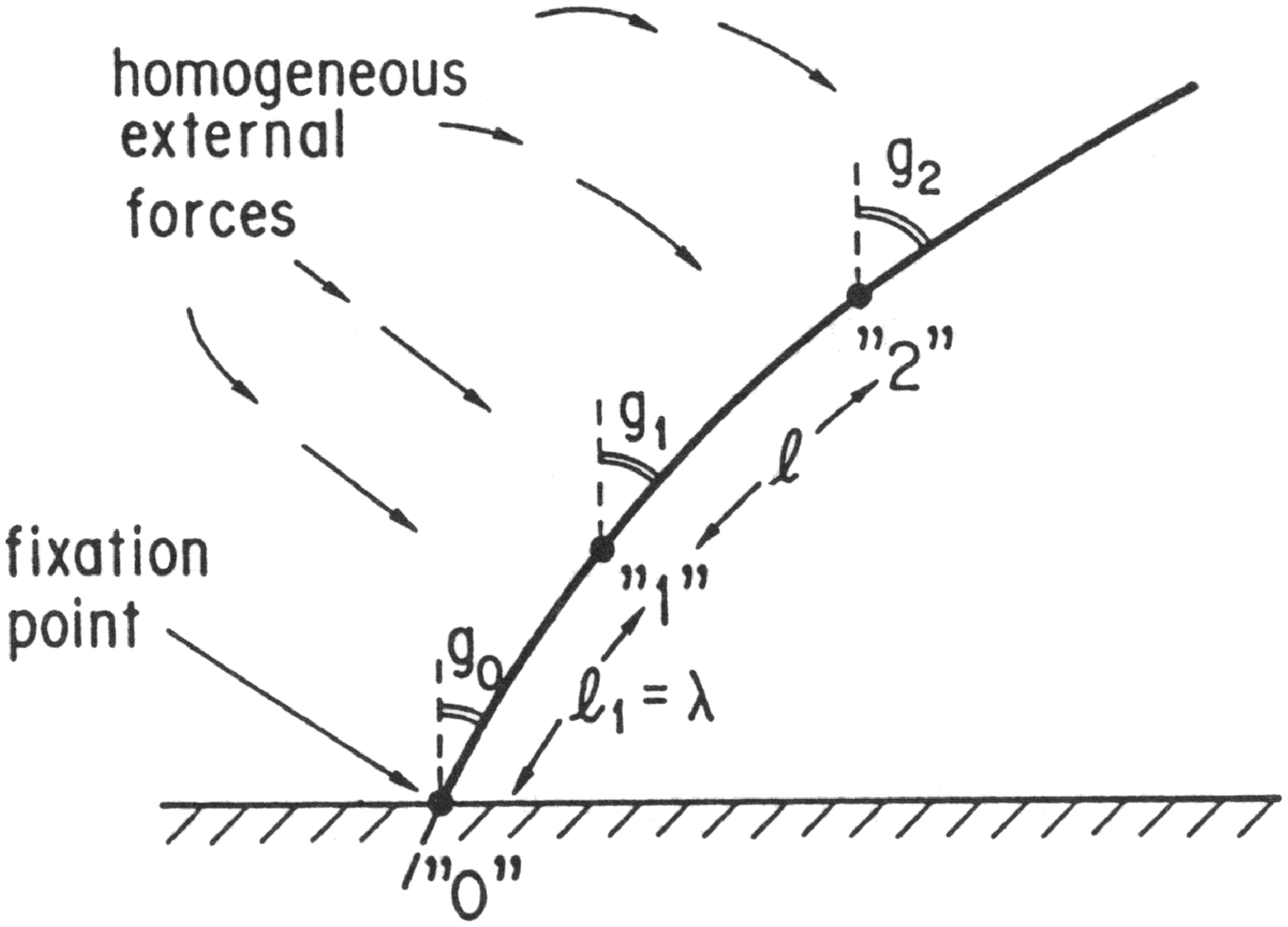}~~~\includegraphics[height=0.34\textwidth]{
  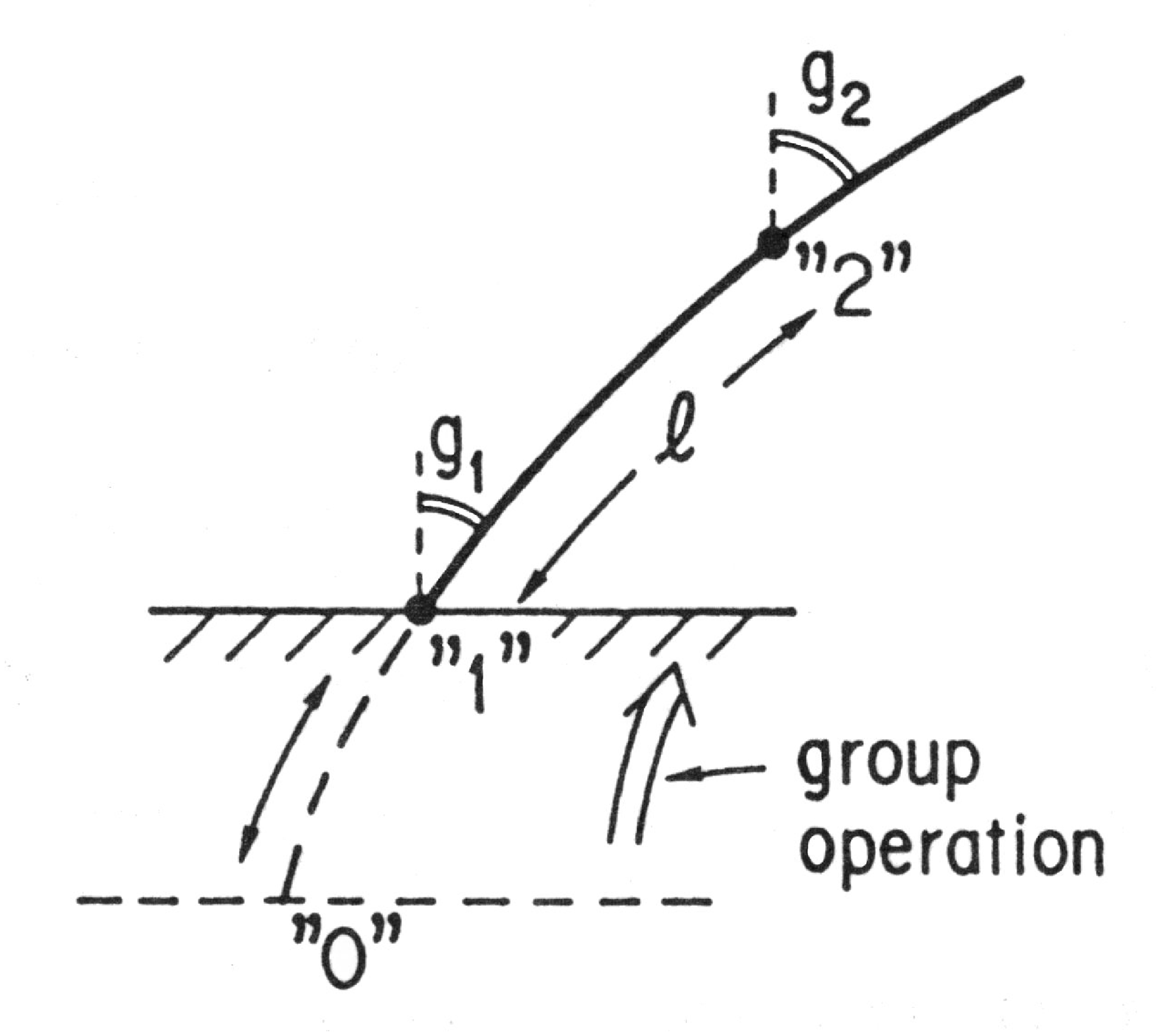}}
  \vspace{-1mm}
   \caption{\small Elastic rod model.\label{fig:ERod}}
\end{figure}

The  symmetry of the FSS group transformations
can be `discovered' in different  fields of physics.
As the illustration, we suggest a mechanical example.
Imagine an elastic rod with a fixed point (point "0")
bent by some external force,
e.g.,
gravity or pressure of a moving gas or liquid,
see in the left panel of Fig.\,\ref{fig:ERod}.
The form of the rod can be described
by the angle \myMath{g}
between tangent to the rod and vertical directions
considered as a function of the distance $l$
along the rod from the fixation point --- by the function \myMath{g(l)}.
If the properties of the rod material and exterior forces
are homogeneous along its length
(independent of \myMath{l}),
then \myMath{g(l)} can  be expressed as function
\myMath{G(l,g_0)},
depending also on \myMath{g_0}
--- deviation angle at the fixation point
from which distance \myMath{l} is measured.
\myMath{G} can depend on other arguments,
like extra forces
and rod material parameters,
but in this context they are irrelevant.

 Take two arbitrary points on the rod, "1" and "2"
with  \myMath{l_1=\lambda} and \myMath{l_2=\lambda+l}.
The angles \myMath{g_i} at points "0", "1" and "2"
are related via \myMath{G} function:
\begin{eqnarray*}
 g_1=G(\lambda,g_0),~~~
 g_2=G(\lambda+l,g_0)=G(l,g_1)\,.
\end{eqnarray*}
To get the RHS of the second equation,
one has to imagine
that the fixation point now is "1",
as shown in the right panel of Fig.\,\ref{fig:ERod}.
Combining both the equations,
\myMath{g_1=G(\lambda,g_0)}
and
\myMath{\displaystyle g_2=G(\lambda+l,g_0)=G(l,g_1)},
one gets the group composition law
\myMath{\displaystyle G\{\lambda, G(l,g)\} = G(\lambda+l,g)}
equivalent
[via relation \myMath{\overline{g}(x,g)\equiv G(l=\ln x,\,g)}]
to FEq
for invariant coupling
\begin{subequations}
 \begin{eqnarray}
  \label{eq:RG-FE}
   \overline{g}(x,g)= \overline{g}(\tfrac{x}{t},\,\overline{g}(t,g))\,.
 \end{eqnarray}
first devised in~\cite{BS56}.
It can be represented in two infinitesimal forms:
The nonlinear differential equation (DEq)
\begin{eqnarray}
 \label{eq:RG-DE}
  x\,\frac{\partial \overline{g}(x,g)}{\partial x}
  = \beta(\overline{g}(x,g)),\quad \mbox{with}
  \quad
  \beta(g)= t\,\frac{\partial\overline{g}(t,g)}
  {\partial\,t}\Big|_{t=1}\,,
\end{eqnarray}
and the linear partial one (PDEq)
\begin{eqnarray}
 \label{eq:RG-PDE}
  \left[x\,\frac{\partial}{\partial x}
       -\beta(g)\,\frac{\partial}{\partial g}
  \right]\,\overline{g}(x,g)=0\,.
\end{eqnarray}
In the QFT jargon, the beta-function
$\beta(g)\,$ is known as the RG generator. \\
Note that each of the two DEqs
is equivalent
to the functional one (\ref{eq:RG-FE})
provided the normalization condition
\begin{eqnarray}
 \label{eq:qft.RG-norm}
  \overline{g}(1,g)=g
\end{eqnarray}
\end{subequations}
is satisfied.
The general solution of the last FEq
can be found from the relation
\begin{eqnarray}
 \label{eq:RGFEq-sol}
  \Phi(\overline{g}) - \Phi(g)
  \equiv \int_g^{\overline{g}}\,
    \frac{d\,\gamma}{\beta(\gamma)}
  = \ln x\,.
\end{eqnarray}

\subsection{RG effective coupling in QFT}
 To illustrate the power of RG-invariance, on
 the one hand, and the effectiveness of its
 differential formulation, consider now an
 illuminating example of the effective coupling
 $\overline{g}$ UV asymptotics. In the common
 perturbation theory (PT), one has \vspace{-5mm}

\begin{subequations}
\begin{eqnarray}
 \label{eq:g.PT.1L}
  \overline{g}_\text{PT}^{[1]}(x;g)
   = g + g^2\beta_0\,\ln x\,,
\end{eqnarray}
 --- the so called one-loop UV logarithm.
Here, as well as in Eqs.\,(\ref{eq:RG-FE})--(\ref{eq:RG-PDE}),
$$x=Q^2/\mu^2\,; \quad g=g_{\mu}= \overline{g}(x;g)\,;
 \quad Q^2=\mathbf Q^2-Q_0^2 > 0\,.$$

Evidently,
expression (\ref{eq:g.PT.1L}) is not RG-invariant.
Indeed, substituting it into the functional equation (\ref{eq:RG.FEq})
one obtains the discrepancy
\begin{eqnarray*}
 \Delta_\text{\,discr}
  \left[\overline{g}_\text{PT}^{[1]}
  \right]
  &\equiv&
   \overline{g}_\text{PT}^{[1]}(x;g)
   - \overline{g}_\text{PT}^{[1]}
      \left(\tfrac{x}{t},\overline{g}_\text{PT}^{[1]}(t;g)\right)
 \\
  &=&
  \left[g + g^2\beta_0\,\ln x\right]
 - \left[g + g^2\beta_0\,\ln x
           + 2g^3\beta_0^2 \ln t\,\ln(x/t)
   \right]\neq 0
\end{eqnarray*}
 --- error of the \myMath{g^3}-order
that can be killed
by adding to starting approximation (\ref{eq:g.PT.1L})
the next-order term \myMath{g^3\,\beta_0^2\,\ln^2x}
etc.
The final result of this iterative restoring
is the famous sum of geometric progression
\begin{eqnarray}
 \label{eq:g.RG.1L}
  \overline{g}_\text{RG}^\text{[1]}(x;g)
   = g\sum_{k\geq0}(g\,\beta_0\,\ln x)^k
   = \frac{g}{1-g\,\beta_0\,\ln x}
\end{eqnarray}
\end{subequations}
which sums up the leading order (LO) logarithms
$g(g\,\ln x)^k\,.$
On the other hand,
this solution can be immediately obtained
by analytical means via the first
differential RG Eq.(\ref{eq:RG-DE})
with
\myMath{\beta(g)=\beta_0\,g^2},
obtained from
Eq.\,(\ref{eq:g.PT.1L}).

Starting with the two-loop perturbative UV
asymptotics
\begin{subequations}
\begin{eqnarray}
 \label{eq:PT.2L}
  \overline{g}_\text{PT}^{[2]}(x;g)= g +g^2\beta_0\ln x
   + g^3\left(\beta_0^2\,(\ln x)^2+\beta_1\ln x \right)
\end{eqnarray}
one gets
(again in few lines of calculations by RG technique)
the Next-to-Leading-Order (NLO)
approximate (at \myMath{\ln x\gg1}) result
\begin{eqnarray}
 \label{eq:PT.iter.2L}
  \overline{g}_\text{RG}^\text{[2]}(x;g)
   \simeq \frac{g}{1 - g\,\beta_0\,\ln x
    - g^2\,\beta_1\,\ln(\ln x)}\,.
\end{eqnarray}
\end{subequations}

\textbf{Illustration in QED.} \
In QED, due to gauge invariance (Ward identities),
the RG-invariant coupling
reduces
(see, pioneer review paper \cite{BS56}
 and references therein as well as
 Sect. 48.1 in monograph \cite{BSvvtkp})
to the transverse amplitude
of the dressed photon propagator
\begin{eqnarray}
 \label{eq:al-qed.def}
 \overline{\alpha}(Q^2,\alpha)
 = \alpha\,d_{tr}(Q^2,\alpha)\,.
\end{eqnarray}
Originally,
it was introduced there as a function
describing the RG transformation
of a PT expansion parameter.
In a proper QED
(quantum electrodynamics of electrons, positrons and photons)
its one-loop expression reads
\vspace{-5mm}
\begin{eqnarray}
 \label{eq:PT.qed1L}
  \overline{\alpha}_\text{PT}^{[1]}(x;\alpha_\mu)
  = \alpha_\mu
  + \frac{\alpha^2_\mu}{3\pi}\,\ln x\,;
\quad
  x = \ln(Q^2/\mu^2)\,.
\end{eqnarray}

\begin{figure}[hb]\vspace{-3mm}
 \centerline{\includegraphics[width=0.55\textwidth]{
  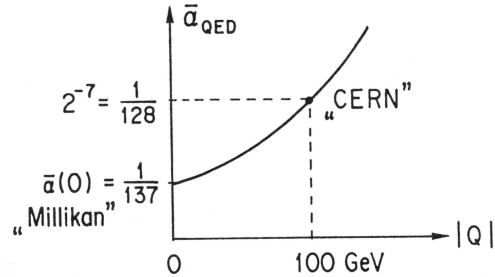}}
  \vspace{-1mm}
   \caption{\small Artistic view on experimental
 verification of the $\overline{\alpha_\text{QED}}\,$
 ``running'' (solid line). \label{fig:QED.alpha}}
\end{figure}
 In a sense,
it is a generalization of the electron effective charge\footnote{Its
 Fourier image is the charge $Q(r)$ of point electron
 screened by quantum vacuum fluctuations.}
 introduced first by Dirac~\cite{Dirac33}:
\vspace{-2mm}
\begin{subequations}
 \begin{eqnarray}
  \label{eq:e(Q)}
   e(Q)\simeq e\,\left[1+\frac{e^2}{12\,\pi^2}\,
  \ln\left(\frac{Q}{m_e}\right)+\ldots \right]\,.
\end{eqnarray}
After the RG machinery, (\ref{eq:PT.qed1L}) takes
the invariant form
\begin{eqnarray}
 \label{eq:RGal.QED}
  \overline{\alpha}_\text{QED}(Q^2)
   \simeq \frac{\alpha_\mu}{1 -
    \left(\alpha_\mu/(3\pi)\right)\,
     \ln\left(Q^2/\mu^2\right)}\,.
\end{eqnarray}
\end{subequations}
Its contemporary Standard Model analog
was checked experimentally
about a decade ago at LEP,
the \myMath{e^+e^-}-collider at CERN,
see results in Fig.\,\ref{fig:QED.alpha}.

Note also that the Bogoliubov RG
was initially devised
in a mass-dependent form~\cite{BS56}.
Starting, e.g., with the mass-dependent one-loop PT
\vspace{-5mm}
\begin{subequations}
\begin{eqnarray} \label{eq:PT.1L.xy}
  \overline{g}^{[1]}_\text{PT}(x,y;g)  = g
   + g^2\left[I_1\left(x/y\right)
      -I_1\left(1/y\right)\right]
\end{eqnarray}
 with $x=q^2/\mu^2$ and $y=m^2/\mu^2\,,$ one gets
\begin{eqnarray} \label{eq:RG.1L.xy}
  \overline{g}_\text{RG}^\text{[1]}(x,y;g) =
   \frac{g}{1-g\,\left[I_1(x/y)-I_1(1/y)\right]}
\end{eqnarray}
\end{subequations}
and, similarly at the two-loop level\cite{Shirkov92}.
This \textit{massive} RG provides us with means to make
an accurate matching across the heavy-quark thresholds,
both in QCD and
in Grand Unification.

\subsection{Analyticity from causality in  QFT}
 Turn now to the \textit{Dispersion Relation Method}
that relates causality in space-time with analyticity
in kinematic (energy, momentum transfer) variables.

  To illustrate the main idea,
consider the
Fourier image \vspace{-1mm}
\begin{eqnarray} \label{eq:F(E)}
 F(E)= \int\limits_{-\infty}^{\infty}
  e^{itE}A(t)\,dt
\end{eqnarray}
of the forward scattering amplitude $A(t)$,
being subdued to the nonrelati\-vistic
\textit{causality condition}:
\begin{eqnarray}
 \label{eq:NRCC}
  A(t)=0\quad\text{at}\quad t<0\,.
\end{eqnarray}
In this case,
\myMath{F(E)} can be analytically continued from real \myMath{E}
values to the upper half of the complex plane
\myMath{Q\to z=E+i\xi;\,\,\xi=\text{Im}\,z>0},
since the factor \myMath{e^{-t\xi}}
in the integrand provides convergence of the integral for \myMath{F(z)}.
Using then the Cauchy theorem with integration contour $\Gamma^+$ in the upper half-plane
and \myMath{z} inside \myMath{\Gamma^+}
\vspace{-2mm}
 \[F(z)=\frac{1}{2\pi\,i}
  \oint_{\Gamma^+}\frac{F(z')}{z'-z}dz'
  \]
one can get \textit{Dispersion Relation}
for the forward scattering amplitude
\begin{eqnarray}
 \label{eq:DR.NonRel}
  \textbf{Re}\,F(E)
  = \frac{1}{\pi}\mathcal P\!\!\int\limits_{\!\!-\infty}^{\,\,\infty}
     \frac{\textbf{Im} F(E')}{E'-E}\,dE'
  = \mathcal P\!\!\int \limits_{\!\!m}^{\,\,\infty}
     \frac{k\sigma(E')}{E'-E}\,dE'\,,
\end{eqnarray}
relating two observable functions.
In obtaining this non-subtracted dispersion relation
we tacitly assumed ``good'' $F(z)$ asymptotic behavior
and used  \textit{Optical Theorem}
\myMath{\textbf{Im}F(E)= k\,\sigma(E)}.
 In a more realistic case, one starts with relativistic
 causality and adds symmetry crossing property of the
 forward scattering amplitude.

  Remind that instead of the Pauli--Jordan commutator
\begin{subequations}
\begin{eqnarray} \label{eq:PWComm}
  D(x-y)=\frac{1}{i}\,\left<0\left|\,\left[
  \phi(x)\,,\,\phi(y)\right]\right|0\right>,
\end{eqnarray}
(vanishing outside the light cone at
$(x-y)^2=(x_0-y_0)^2-(\text{x--y})^2<0$
and involved in relativistic-invariant quantization)
in construction of matrix elements
and observables
one needs the Stueckelberg--Feynman causal propagator
\vspace{-2mm}
\begin{eqnarray}
 \label{eq:Dc}
  D_\text{c}(x-y)=D_\text{F}(x-y)=\frac{1}{i}\,
  \left<0\left|T\,\left[\phi(x)\,\phi(y)
  \right]\right|0\right>\,,
\end{eqnarray}
being the vacuum expectation value of
the time-ordered product.
Just this function (and its derivatives)
enters into Feynman rules. \\
At the same time, in the Schwinger--Dyson equations
one deals with its ``dressed'' version
which includes
radiative corrections
\begin{eqnarray} \label{eq:Ddres}
  D_\text{dress}(x;g)
   &=& \frac{1}{i\,S_0}\,
      \left<0\left|T\,\left[\phi(x)\,\phi(0)\,S(g)\right]
            \right|0      \right>\,,
\end{eqnarray}
from the scattering matrix
\vspace{-1mm}
\begin{eqnarray} \label{eq:Smatrix}
  S(g) &=&1 +g\,S_1 +g^2\,S_2 +\ldots\nonumber
\end{eqnarray}
 This dressed causal propagator
can be represented in the K\"allen--Lehmann
spectral form\footnote{
Here we give its simplest, nonsubtracted version
that is valid for the appropriately decreasing
spectral density behaving as
$\rho(\sigma,g) \lesssim 1/\ln^2\sigma\,$).}
\vspace{-1mm}
\begin{eqnarray} \label{eq:Ddress.KL}
 D_\text{dress}(q^2, g)=\frac{1}{m^2-q^2-
 \text{i}\varepsilon}+\frac{1}{\pi}\,\int\limits_{\!
 \!m_1}^{\,\,\infty}\frac{\rho(\sigma,g)\,d\sigma}
            {\sigma-q^2-\text{i}\varepsilon}
\end{eqnarray}
\end{subequations}

This representation, in a sense,
resembles the forward dispersion relation
(\ref{eq:DR.NonRel}).
Here, quite similarly,
it defines the function
$\,D(z)\equiv D_\text{dress}(z, g)\,$
analytic in the whole complex plane $z$
except the pole and the cut on part of the real axis.
In its proof, as well,
the relativistic generalization of the causality condition
(\ref{eq:NRCC}) is used.

 In what follows, we have to combine two nonperturbative methods
--- the RG and the Dispersion Relation ones.

\subsection{RG and causality}
 \label{sec:1-5}             
 Return to the QED invariant coupling.
According to Eq.\,(\ref{eq:al-qed.def}), it is proportional
to the transverse amplitude of the dressed photon propagator.
As it was first shown by K\"allen~\cite{Kall52},
the latter amplitude can be represented
in the form congeneric to Eq.\,(\ref{eq:Ddress.KL}).
Hence, the same presentation
\vspace*{-1mm}
\begin{eqnarray} \label{eq: spectr-albar}
 \overline{\alpha}(Q^2,\alpha)\,=\frac{1}{\pi}\,
  \int\limits_{\!\!0}^{\,\,\infty}
   \frac{\rho(\sigma,\alpha)\,d\sigma}
        {\sigma+Q^2-\text{i}\varepsilon}
\end{eqnarray}
has to be valid for the QED effective coupling.
That is,
the function $\overline{\alpha}(z,\alpha)$ should be analytic
in the duly cut complex $z$ plane.
This is true, term-by-term, for its PT expansion,
like Eq.\,(\ref{eq:PT.qed1L}).
However, it is not true for its RG-invariant counterpart,
Eq.\,(\ref{eq:RGal.QED}),
as it contains parasitic singularity
outside the allowed cut ---
the so-called ghost or Landau pole at
\begin{eqnarray}
 \label{eq: Dau-pole}
  Q^2=Q_*^2=\mu^2\,e^{\,3\pi/\alpha_\mu}\,.
\end{eqnarray}
An elegant solution to resolve the issue
was proposed by Bogoliubov et al.~\cite{BLS60}.
Omitting the details we mention here the main recipe:
\begin{quote}\textit{
 To bring a RG-invariant but singular (containing
 extra pole) expression in the proper
 $Q^2$-analytic form, one has to use
 the spectral representation (\ref{eq: spectr-albar})
 with the spectral density $\rho$ defined from the
 related PT input, Eq.\,(\ref{eq:RGal.QED}).}
\end{quote}
The resulting \textit{analyticized} expression
is
\begin{eqnarray} \label{eq:anal-albar}
 \overline{\alpha}_\text{an}(x,\alpha_\mu)
  = \frac{\alpha_\mu}
         {1 - \alpha_\mu/(3\pi)\,\ln x}\,
  + \frac{3\pi}{x-x_*}\,;
\quad x_*\equiv e^{\,3\pi/\alpha_\mu}\,.
\end{eqnarray}
The second term in the RHS is invisible
in the PT expansion.
Remarkably, it contains an essential singularity
at $\alpha=0\,$ of a proper type.

\subsection{QCD effective coupling}
 \label{sec:tab-PT}
In QCD, at the one-loop, LO approximation one has
\begin{eqnarray}
 \label{eq:as.1L}
  \alpha_s^{(1)}(Q)
  = \frac{\alpha_s(\mu)}{1+\alpha_s(\mu)
     \beta_0\,\ln(Q^2/\mu^2)}
  = \frac{1}{\beta_0\,L_Q},\,\,
  L_Q = \ln\left(\frac{Q^2}{\Lambda^2}\right)\,.
\end{eqnarray}
The NLO, or two-loop expression in
the Denominator Representation
contains log-of-log dependence
(just like in eq.(\ref{eq:PT.iter.2L}) for \myMath{L_Q\gg1}):
\begin{eqnarray}
 \label{eq:as.2L.Den}
  \alpha_s^{(2)}(Q)
  \simeq \frac{1}{\beta_0L_Q+(\beta_1/\beta_0)\ln L_Q}\,;
 \quad \beta_{0,1}\sim 1\,.
\end{eqnarray}
Note that the famous Asymptotic Freedom UV behavior
$\alpha_s(Q)\sim 1/L_Q $ is correct
already in the LO approximation.
The QCD scale \myMath{\Lambda} turns out
to be numerically close
to the confinement scale
\myMath{\Lambda\sim 300-400\,\text{MeV}
\simeq 2\,m_\pi};
that is \myMath{R_\Lambda \sim 10^{-13}}\,cm.

\begin{figure}[h!]\vspace{-3mm}
 \centerline{\includegraphics[width=0.45\textwidth,height=0.4\textwidth]
  {
   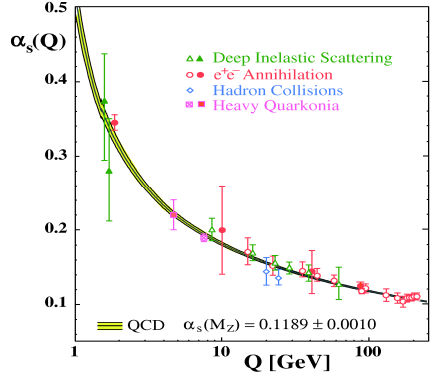}}\vspace{-1mm}
   \caption{\small Triumph of the NLO pQCD from
   200 up to $3-5$ GeV (taken from~\cite{Bethke07}).
   \label{fig:QCD.alpha}}
\end{figure}

The QCD final product is formulae for hadronic
observables.
Some of them, e.\,g.,
the ratio of inclusive cross-sections for \myMath{e^+e^-} annihilation,
can be directly expressed\vspace{-1mm}
\begin{subequations}
\begin{eqnarray}
 \label{eq:R.Def}
  R_{e^+e^-}(s)
  = \frac{\sigma_{e^+e^-\to\text{hadrons}}(s)}
         {\sigma_{e^+e^-\to\mu^+\mu^-}(s)}
  = R\left(s;\alpha_s\right)
\end{eqnarray}
in terms of the QCD notions.
As an RG-invariant
this ratio
\textit{should depend on the QCD coupling \
\myMath{\alpha_s(s)} \ only}!
Perturbatively, it is power functional expansion
\begin{eqnarray}
 \label{eq:R.PT}
  R(s;\alpha_s)= R_\text{inv}(\alpha_s(s))
  = 1 + r_1\,\alpha_s(s)
      + r_2\,\alpha_s^2(s)
      + \text{O}\left(\alpha_s^3(s)\right)\,.
\end{eqnarray}
\end{subequations}
Remarkably that,
according to the 2007 Bethke review~\cite{Bethke07},
above a few GeV the two-loop pQCD nicely correlates
all \myMath{\alpha_s(Q^2)} data,
see Fig.\,\ref{fig:QCD.alpha}.

 {\small\begin{table}[b!]  \centering
 \begin{tabular}{|l||r||c|c|c|c|c|}\hline
 {\slshape\phantom{a}Observable$\vphantom{^|_|}$}
     &Scale~~~&1-loop &2-loop &3-loop &4-loop & \myMath{\Delta_\text{exp}}\\ \hline\hline
 \myMath{R_{e^+e^-\to \text{hadrons}}}
     & 10 GeV & 92\%  & 7.6\% & 1.0\% &\myMath{-0.6\%\vphantom{^|_|}}
                                              & 12--30\% \\ \hline
 \myMath{R_\tau} in \myMath{\tau}-decay
     & 2 GeV  & 51\%  & 27\%  & 14\%  & \myMath{8\%\vphantom{^|_|}}
                                              & 5\% \\ \hline
 {\small Bjorken SR}
     & 2 GeV  & 56\%  & 21\%  & 12\%  & \myMath{11\%\vphantom{^|_|}}
                                              & 6\% \\ \hline
 \end{tabular}
 \caption{\small Relative size of 1-, 2-, 3-, and 4-loop
 contributions to observables.
 Two last lines are taken from~\cite{Khan11}.
 Three-loop estimations can be found in~\cite{SS06}.
 \label{tab:PT}}\end{table} } 

However, data below 5~GeV are not good enough:
One can see from Table~\ref{tab:PT}
that while the situation with higher-loop corrections
for the $R_{e^+e^-}$ ratio is rather good (the first line),
the Bjorken Sum Rule (SR) case (the third line)
delivers to us a signal of possible blow-up
(like in Table~\ref{tab:I(g)})
of the related asymptotic perturbative series
at $Q\lesssim 2-3$ GeV.
Just in this region the PDG canonized explicit expression
for \myMath{\alpha_s(Q^2)} starts to grow
sharply as even the one-loop expression for effective
 QCD coupling\vspace*{-1mm}
\begin{eqnarray}
 \label{eq:as.sing}
  \alpha_s^{(1)}(Q)
  = \frac{1}{\beta_0\,\ln(Q^2/\Lambda^2)}
 \sim \frac{1}{\beta_0\,(Q^2-\Lambda^2)}
 \quad\text{at}\quad Q^2\sim\Lambda^2
\end{eqnarray}\vspace*{-1mm}
has \textit{unphysical} (Landau) singularity.
Just this singularity at \myMath{|Q|=\Lambda \sim 350} MeV
prevents from analyzing data
by pQCD in the low-energy physical region
below few GeV.

Meanwhile, all nonperturbative lattice-QCD simulations
testify to regularity of \myMath{\alpha_s(Q)} behavior
in the \myMath{Q^2\sim\Lambda^2} region
--- see fresh overviews \cite{Shi02lat}.
In what follows,
we shall deal with ghost-free analytic QCD couplings
(and their powers)
taken from the APT and its generalization
which are regular in the low-energy region
and in the high-energy one
coincide with the common $\alpha_s(Q^2)$.

\section{Analytic Perturbative Theory (APT) in QCD}
 \subsection{APT: Historic Preamble \label{2-1}}
 As it has been mentioned above in Sect.~\ref{sec:1-5},
the analytization recipe is a common product
of two nonperturbative methods
--- the RG and DR ones.
Originally, it was formulated~\cite{BLS60}
for QED in the Euclidean region.
Then in 1982 Radyushkin
and Krasnikov\&Pivovarov \cite{RKP82}
using the dispersion technique suggested regular
(for $s\geq \Lambda^2$)
QCD running couplings in the Minkowski region,
namely $\pi^{-1}\arctan(\pi/L_s)$ and
\myMath{\Phi(\alpha_s^n[L_s]),\,L_s=\ln(s/\Lambda^2)},
with indication of the need to use
the nonpower expansion in \myMath{\Phi(\alpha_s^n[L_s])}
instead of the power one.
The real proliferation of this technique
into QCD was initiated
by Igor Solovtsov
and his co-authors in the mid-90s~\cite{APTbasic}.
In these pioneer papers, the ghost-free expressions
for the RG-invariant QCD couplings
in both the energy-like Minkowskian region
and the momen\-tum-transfer Euclidean one
were obtained.
Quite soon this construction
was developed~\cite{MSSDVBRS}
in a closed scheme.\footnote{
The essential point was discovering the necessity
of nonpower type of functional expansions~\cite{DVShi99}
which are the only compatible ones with linear integral transformations relating Euclidean, Minkowski and space
 position pictures.}
See the history details in the recent review paper~\cite{SS06}.
The whole construction is known since then
as the Analytic Perturbation Theory.\footnote{Close results
were partly obtained by Simonov
using the background perturbation theory~\cite{Sim01}.}

The next step, made by Bakulev, Mikhailov, and Stefanis,
generalizes the APT
by including fractional powers of coupling,
as well as products of coupling powers and logarithms~\cite{BMS-FAPT}
and for this reason, it was named the \textit{Fractional APT}.
At the same time, it appears possible to sum up
nonpower series in the (F)APT~\cite{MS04,BMS-Resum}.

 \subsection{Basics of APT}
In the standard pQCD,
as we described in Sect.~1,
the one-loop RG equation
for the effective coupling
$\alpha_s(Q^2)=a[L]/\beta_f$
with $L=\ln(Q^2/\Lambda^2)$
and
$\beta_f=\beta_0(N_f)=(11-2N_f/3)/(4\pi)$
generates the pole singularity,
$a[L]=1/L$.

Due to this pole, in pQCD
the problem arises:
How to go to the Minkowski region?
Quantities in the Minkowski region are usually represented
by contour integrals of the type \myMath{\oint f(z)D(z)dz}.
In the integrands one uses \myMath{D(z)=\sum\nolimits_m d_m\alpha_s^m(z)}
and changes the integration contour to \myMath{\Gamma_i}.
This change of the integration contour is legitimate if
\myMath{\displaystyle D(z)f(z)} is analytic inside the circle everywhere.
But \myMath{\alpha_s(z)} and hence \myMath{D(z)f(z)}
have the Landau pole singularity just inside!
\begin{figure}[t!]\vspace{-3mm}
 \centerline{
 \includegraphics[width=0.45\textwidth]{
  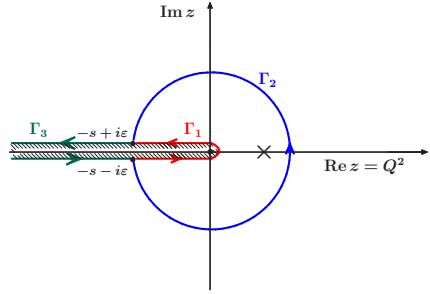}}
  \vspace{-1mm}
   \caption{\small Contours of integration on the road to the Minkowski space.
    \label{fig:Contours}}
\end{figure}
In the APT effective couplings \myMath{\displaystyle \mathcal A_n(z)}
are analytic functions
and this problem does not appear at all!
In~\cite{BMS-Resum} (2010)
the equivalence of the Contour-Improved-PT (CIPT) approach to the (F)APT
for quantities like \myMath{R(s)}
was proved,
which can be symbolically expressed
as \vspace{-1mm}
\begin{center}
 \framebox{\textbf{CIPT}\myMath{\displaystyle\left\{\oint_{\Gamma_2}\frac{D(z)dz}{z}\right\}
  =} \textbf{(F)APT}\myMath{\displaystyle\left\{\oint_{\Gamma_3}\frac{D(z)dz}{z}\right\}}\,.}
\end{center} \vspace{-1mm}

By the analytization in the APT for an observable $f(Q^2)$
we mean the K\"allen--Lehmann representation (\ref{eq:Ddress.KL})
\begin{eqnarray}
 \label{eq:An.SD}
  \left[f(Q^2)\right]_\text{an}
   = \int_0^{\infty}\!
      \frac{\rho_f(\sigma)}
         {\sigma+Q^2-i\epsilon}\,
       d\sigma
\end{eqnarray}
with $\displaystyle\rho_f(\sigma)=\frac{1}{\pi}\,\textbf{Im}\,\big[f(-\sigma)\big]$.
Then in the one-loop approximation $\overline{\rho}_1(\sigma)=1/\sqrt{L_\sigma^2+\pi^2}$
and\footnote{
   We use the notation $f(Q^2)$ and $f[L]$ in order to specify the arguments we mean ---
   squared momentum $Q^2$ or its logarithm $L=\ln(Q^2/\Lambda^2)$,
   that is $f[L]=f(\Lambda^2\cdot e^L)$ and $\Lambda^2$
   is usually referred to the $N_f=3$ region.
   Note that here we introduced the notation
   \myMath{\overline{\mathcal A}_n[L]} and
   \myMath{\overline{\mathfrak A}_n[L]}
   in order to distinguish analytic images of normalized coupling powers,
   \myMath{a^n(Q^2)}, from the corresponding images of \myMath{\alpha_s^n(Q^2)}
   --- for the latters we use the standard notation
   \myMath{\mathcal A_n[L]} and \myMath{\mathfrak A_n[L]}.
   This notation is different from that used in~\cite{BMS-FAPT,BMS-Resum}.}
\begin{subequations}
 \label{eq:A.U}
 \begin{eqnarray}
  \label{eq:A_1}
 \overline{\mathcal A}_1[L]
  &=& \int_0^{\infty}\!\frac{\overline{\rho}_1(\sigma)}{\sigma+Q^2}\,d\sigma\
   =\ \frac{1}{L} - \frac{1}{e^L-1}\,,~\\
 \label{eq:U_1}
 \overline{\mathfrak A}_1[L_s]
  &=& \int_s^{\infty}\!\frac{\overline{\rho}_1(\sigma)}{\sigma}\,d\sigma\
   =\ \frac{1}{\pi}\,\arccos\frac{L_s}{\sqrt{\pi^2+L_s^2}}\,,~
 \end{eqnarray}
whereas analytic images of the higher powers ($n\geq2, n\in\mathbb{N}$) are:
\begin{eqnarray}
 \label{eq:recurrence}
 {\overline{\mathcal A}_n[L] \choose \overline{\mathfrak A}_n[L_s]}
  &=& \frac{1}{(n-1)!}\left[-\frac{d}{d L}\right]^{n-1}
      {\overline{\mathcal A}_{1}[L] \choose \overline{\mathfrak A}_{1}[L_s]}\,.
\end{eqnarray}
\end{subequations}
Note that at \myMath{L\gg1} the pole remover
\myMath{\sim e^{-L}\approx e^{-1/a}}.
In other words,
K\"allen--Lehmann analyticity
in the \myMath{Q^2} plane
generates nonperturbative
\myMath{e^{-1/\alpha_s}} correction.
This correction guarantees the absence of spurious Lan\-dau-pole singularity
and
ensures the correspondence
with PT \myMath{\alpha_s(Q^2)} at \myMath{Q^2\gg 1\text{~GeV}^2}.

\begin{figure}[h!]\vspace{-1mm}
 \centerline{\includegraphics[width=0.47\textwidth]{
 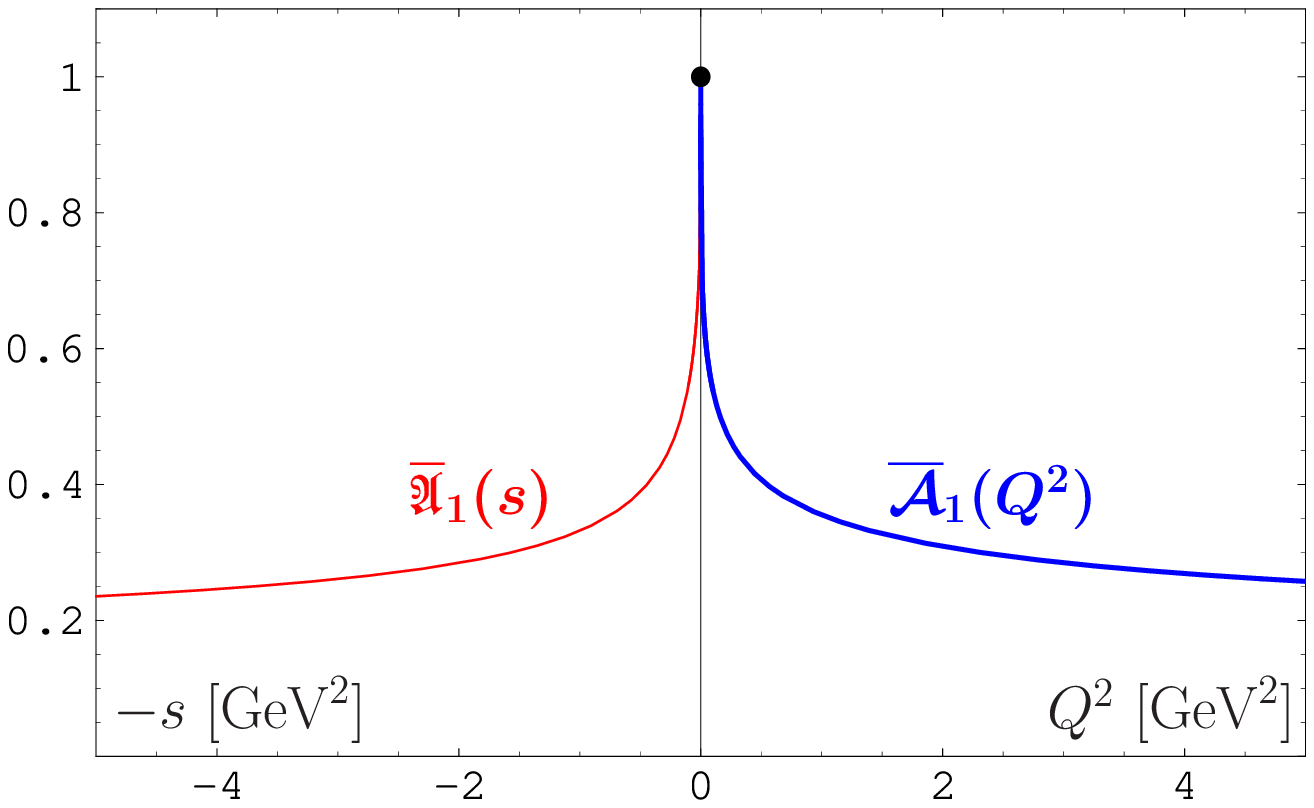}~\includegraphics[width=0.48\textwidth]{
 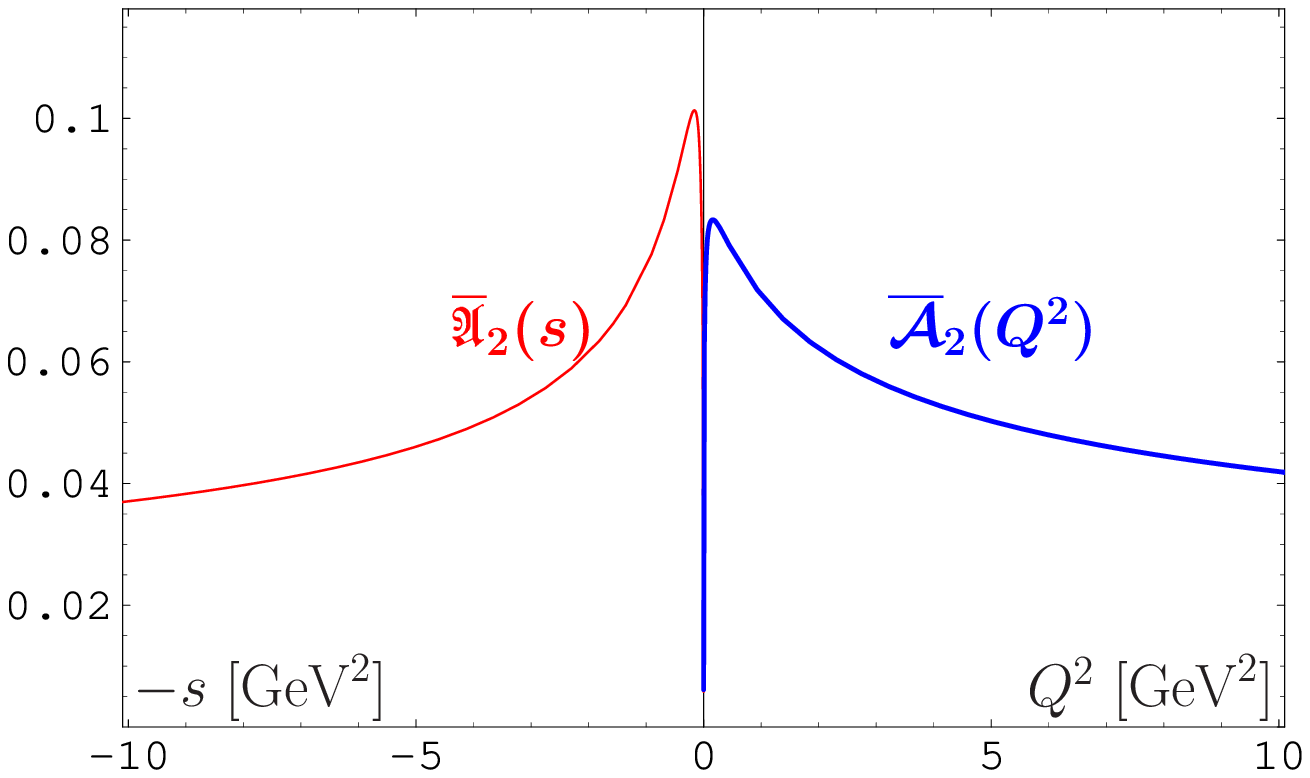}
   \vspace{-1mm}}
    \caption{\small Distorting Mirror for analytic couplings in the Minkowski and
     Euclidean regions. Left panel: for \myMath{\overline{\mathfrak A}_{1}(s)}
     and \myMath{\overline{\mathcal A}_{1}(Q^2)}.
     Right panel: for \myMath{\overline{\mathfrak A}_{2}(s)}
     and \myMath{\overline{\mathcal A}_{2}(Q^2)}.\label{fig:DistMirror}}
\end{figure}
In Fig.\,\ref{fig:DistMirror}, we show the so-called {Distorting Mirror}
for analytic couplings in the Minkowski and Euclidean regions:
in the left panel ---
for \myMath{\overline{\mathfrak A}_{1}(s)} and \myMath{\overline{\mathcal A}_{1}(Q^2)},
whereas in the right panel ---
for \myMath{\overline{\mathfrak A}_{2}(s)} and \myMath{\overline{\mathcal A}_{2}(Q^2)}.
We see that in the IR domain one has universal finite IR values
\myMath{\overline{\mathcal A}_{1}(0) = \overline{\mathfrak A}_{1}(0) = 1}.
Moreover, starting from the two-loop level analytic couplings
reveal loop stabilization of IR behavior.
This yields practical loop- and renormaliza\-tion-scheme-independence of
\myMath{\mathcal A_{1}(Q^2)}, \myMath{\mathfrak A_{1}(s)},
and higher expansion functions, for details see~\cite{SS06}.

\begin{figure}[t!]\vspace*{-1mm}
 \centerline{\includegraphics[width=0.45\textwidth,height=0.3\textwidth]{%
  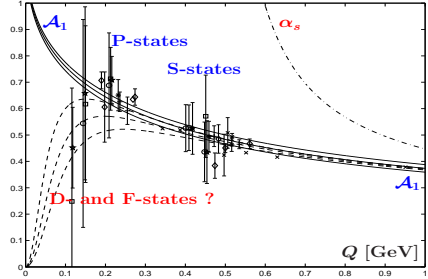}}
  \caption{\small Comparing \myMath{\alpha_s^\text{exp}} from the meson
  spectrum and the three-loop \myMath{\mathcal A_{1}}
   at \myMath{\Lambda^{(3)}_{n_f=3}=(417\pm 42)}\,MeV (solid lines).
   The three-loop \myMath{\alpha_s} is shown as the dot-dashed line.
   \label{fig:Mes.Spectr}}
 \end{figure}
In Fig.\,\ref{fig:Mes.Spectr}, we show the results of the APT application
in meson spectroscopy
obtained in~\cite{BNPSS07}.
We observe from this comparison
that the three-loop \myMath{\alpha_s},
shown as dot-dashed line,
is completely excluded,
whereas the three-loop APT coupling goes
through all the ``experimental'' points
due to well-established $P$- and $S$-wave meson states.

In the APT,
in addition to the regular behavior of couplings,
one has to have nonpower expansions for physical observables.
Indeed, if in the standard pQCD
for an observable \myMath{D} we have\footnote{
Here \myMath{r_3} and higher-order coefficients \myMath{r_n} differ
from \myMath{d_3} and \myMath{d_n}
by the \myMath{\pi^2} terms.}
\begin{subequations}
 \label{eq:Exp.PT}
\begin{eqnarray}
  D_\text{PT}(Q^2)
   &=& d_0
   + d_1\,\alpha_s(Q^2)
   + d_2\,\alpha_s^2(Q^2)
   + d_3\,\alpha_s^3(Q^2)
   + \ldots\,;~~~\\
  R_\text{PT}(s)
   &=& d_0
   + d_1\,\alpha_s(s)
   + d_2\,\alpha_s^2(s)
   + r_3\,\alpha_s^3(s)
   + \ldots\,,~~~
\end{eqnarray}
then in the APT
we should use the nonpower functional expansion
\begin{eqnarray}
 \label{eq:Exp.APT.E}
  \mathcal D_\text{APT}(Q^2)
   &=& d_0
   + d_1\,\mathcal A_1(Q^2)
   + d_2\,\mathcal A_2(Q^2)
   + d_3\,\mathcal A_3(Q^2)
   + \ldots\,;~~~\\
 \label{eq:Exp.APT.M}
  \mathcal R_\text{APT}(s)
   &=& d_0
   + d_1\,\mathfrak A_1(s)
   + d_2\,\mathfrak A_2(s)
   + d_3\,\mathfrak A_3(s)
   + \ldots\,.~~~
\end{eqnarray}
\end{subequations}
This provides
 \begin{itemize}
 \item Better loop convergence (compare Tables~\ref{tab:PT} and \ref{tab:APT}) and practical
  renor\-malization-scheme independence of observables;
  \item Third terms in (\ref{eq:Exp.APT.E}) and (\ref{eq:Exp.APT.M})
   contribute less than 5\%, cf.\,Table~\ref{tab:APT}.
   Again the two-loop (N\myMath{^2}LO) level is sufficient.
 \end{itemize}\vspace*{-1mm}

\begin{table}[h!]
 \centering
  \begin{tabular}{|l||r||c|c|c|c|c|}\hline
 {\slshape\phantom{a}Observable$\vphantom{^|_|}$}
          & Scale & 1-loop & 2-loop & 3-loop &4-loop & \myMath{\Delta_\text{exp}}\\ \hline\hline
 \myMath{R_{e^+e^-\to \text{hadrons}}}
          & \myMath{10} GeV
                  & \myMath{92.2\%}
                           & \myMath{7\%}
                                    & \myMath{0.7\%}
                                             & \myMath{0.1\%\vphantom{^|_|}}
                                                     & 12--30\% \\ \hline
 \myMath{R_\tau} in \myMath{\tau}-decay
          & \myMath{2} GeV
                  & \myMath{90.6\%}
                           & \myMath{8.2\%}
                                    & \myMath{1\%}
                                             & \myMath{0.2\%\vphantom{^|_|}}
                                                     & 5\% \\ \hline
 Bjorken SR
          & \myMath{2} GeV
                  & \myMath{75\%}
                           & \myMath{20.5\%}
                                    & \myMath{4.55\%}
                                             & \myMath{-0.05\%\vphantom{^|_|}}
                                                     & 6\% \\ \hline
 \end{tabular}
 \caption{\small Relative size of 1-, 2-, 3-, and 4-loop
  contributions to observables in the APT (for comparison with PT ---
  see Table~\ref{tab:PT} in Section~\ref{sec:tab-PT}).
  Estimates on \myMath{R_\tau} are taken from~\cite{Mag10}, whereas
  those on Bjorken SR --- from the paper in preparation
  \cite{Khan11} (the 3-loop estimations can be found in~\cite{SS06}).
  \label{tab:APT}}
\end{table}

 \subsection{From APT to FAPT}
At first glance, the APT is a complete theory
providing tools to produce
an analytic answer for any perturbative series in QCD.
However, in 2001 Karanikas and Stefanis~\cite{KS01}
suggested the principle of analytization ``as a whole''
in the $Q^2$ plane for hadronic observables,
calculated perturbatively.
More precisely, they proposed the analytization recipe
for terms like
$\int_{0}^{1}\!dx\!\int_{0}^{1}\!dy\,
  \alpha_\text{s}\left(Q^{2}xy\right) f(x)f(y)$
which can be treated as an effective account
for the logarithmic terms
in the next-to-leading-order
approximation of the pQCD.
Indeed, in the standard pQCD one also has:\\
(i) the factorization procedure in QCD
    that gives rise to the appearance of logarithmic factors of the type:
     $a^\nu[L]\,L$;\\
(ii) the RG evolution
     that generates evolution factors of the type:\\
     $B(Q^2)=\left[Z(Q^2)/Z(\mu^2)\right]$ $B(\mu^2)$
     which reduce in the one-loop approximation to
     $Z(Q^2) \sim a^\nu[L]$ with $\nu=\gamma_0/(2\beta_0)$
     being a fractional number.\\
All that means
that in order to generalize the APT
in the ``analytization as a whole'' direction,
one needs to construct analytic images
of new functions:
$\displaystyle a^\nu,~a^\nu\,L^m, \ldots$\,.
This task was performed in the framework of the so-called FAPT
suggested in~\cite{BMS-FAPT}.
Now we briefly describe this approach.

\begin{figure}[t!]
 \begin{minipage}{\textwidth}
  \centerline{\includegraphics[width=0.49\textwidth]{
  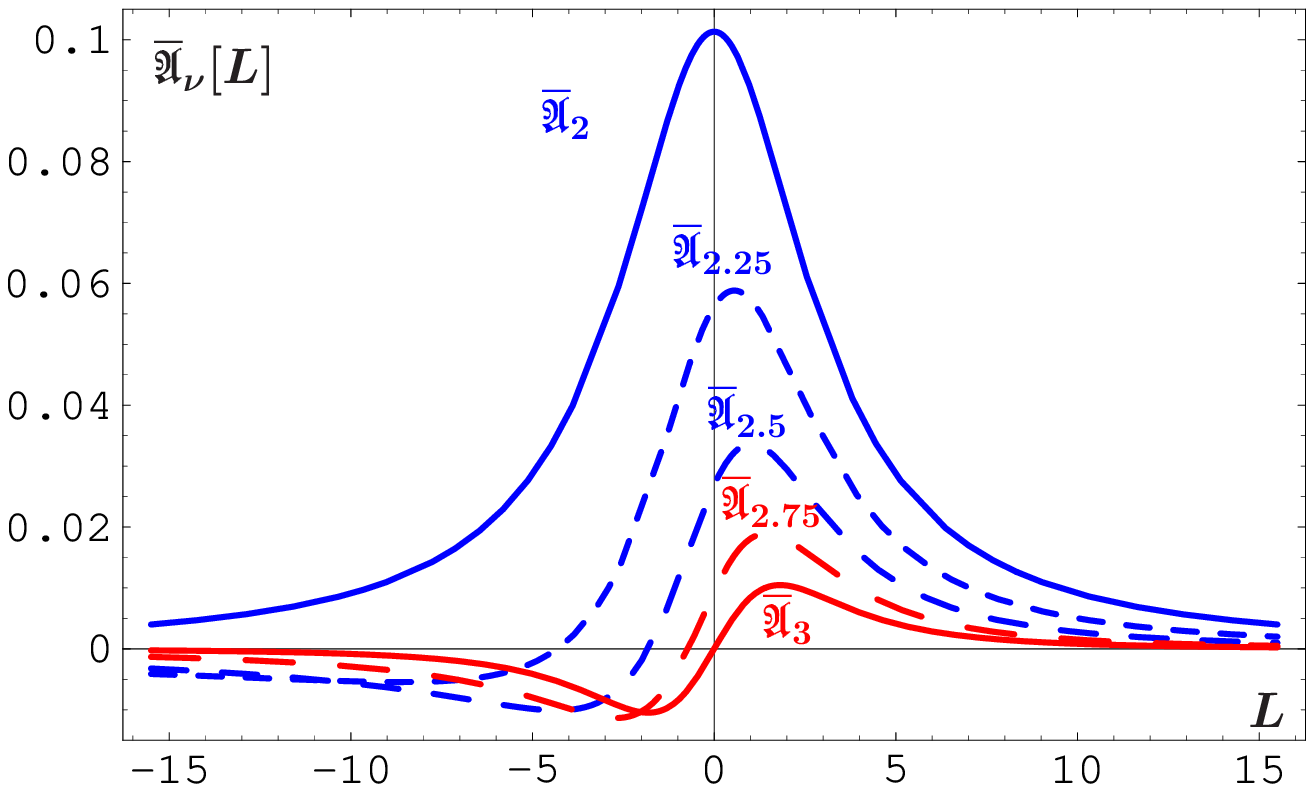}~\includegraphics[width=0.49\textwidth]{
  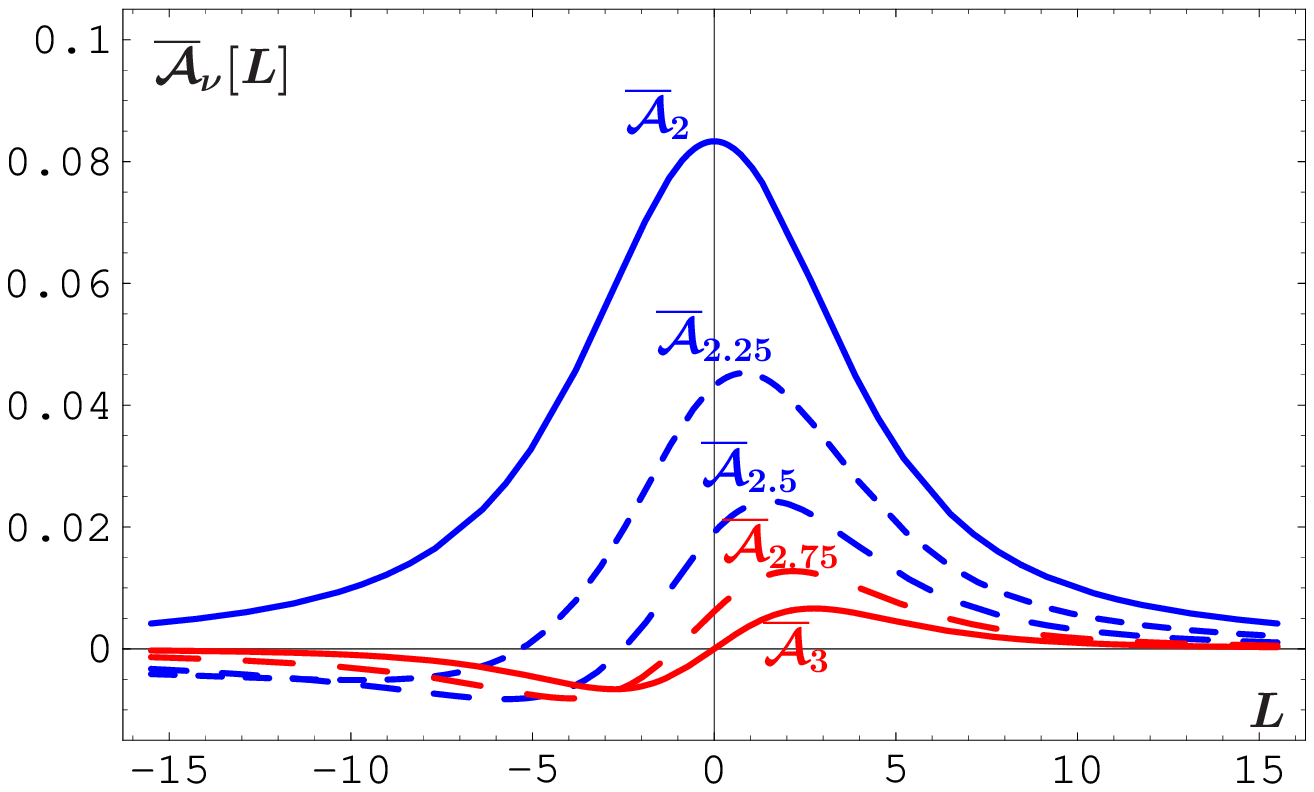}\vspace*{-1mm}}
   \caption{\small Comparing  \myMath{\overline{\mathfrak A}_{\nu}[L]} (left panel) and
    \myMath{\overline{\mathcal A}_{\nu}[L]} (right panel) vs. \myMath{L}
    for fractional \myMath{\nu\in\left[2,3\right]}.
    \label{fig:UCal.ACal.nu}}
\end{minipage}
\end{figure}

In the one-loop approximation,
using recursive relation (\ref{eq:recurrence})
we can obtain explicit expressions for
${\overline{\mathcal A}}_{\nu}[L]$
and ${\overline{\mathfrak A}}_{\nu}[L]$:
\begin{subequations}
\begin{eqnarray}
 \overline{\mathcal A}_{\nu}[L]
 &=&
   \frac{1}{L^\nu}
  - \frac{F(e^{-L},1-\nu)}{\Gamma(\nu)}\,;
 ~
 \\
 \overline{\mathfrak A}_{\nu}[L]
 &=&
   \frac{\text{sin}\left[(\nu -1)\arccos\left(L/\sqrt{\pi^2+L^2}\right)\right]}
         {\pi(\nu -1)\left(\pi^2+L^2\right)^{(\nu-1)/2}}\,.~
\end{eqnarray}
\end{subequations}
Here $F(z,\nu)$ is the reduced Lerch transcendental function
which is an analytic function in $\nu$.
The couplings \myMath{\overline{\mathcal A}_{\nu}[L]}
and \myMath{\overline{\mathfrak A}_{\nu}[L]}
have very interesting properties,
which we discussed extensively in our previous papers~\cite{BMS-FAPT}.

In Fig.\,\ref{fig:UCal.ACal.nu}, we show in comparison
how \myMath{\overline{\mathfrak A}_{\nu}[L]}
and \myMath{\overline{\mathcal A}_{\nu}[L]}
depend on \myMath{L} for fractional values of
\myMath{\nu}:
one more time we observe the same picture of the Distorting Mirror
when comparing the Minkowski (left panel)
and Euclidean (right panel)
regions.

\begin{figure}[h!]
 \begin{minipage}{\textwidth}
  \centerline{\includegraphics[width=0.49\textwidth]{
   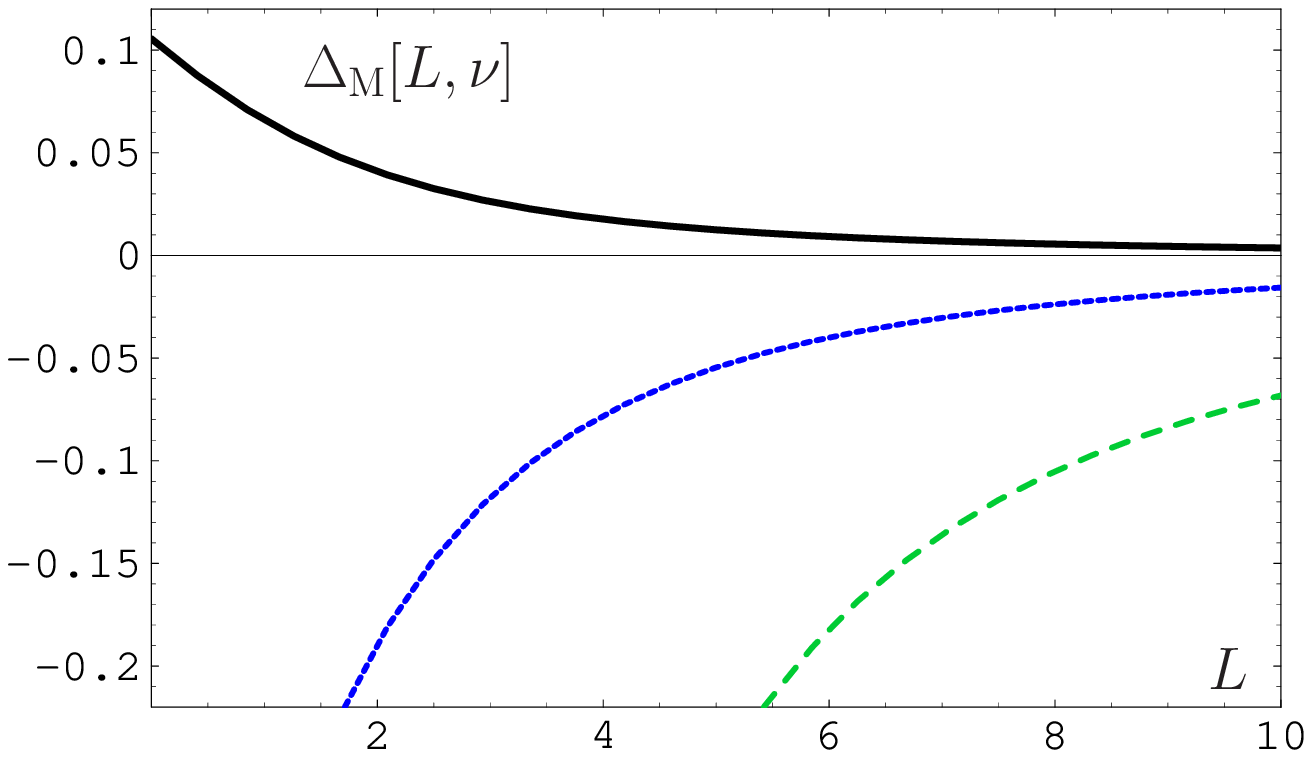}~\includegraphics[width=0.49\textwidth]{
   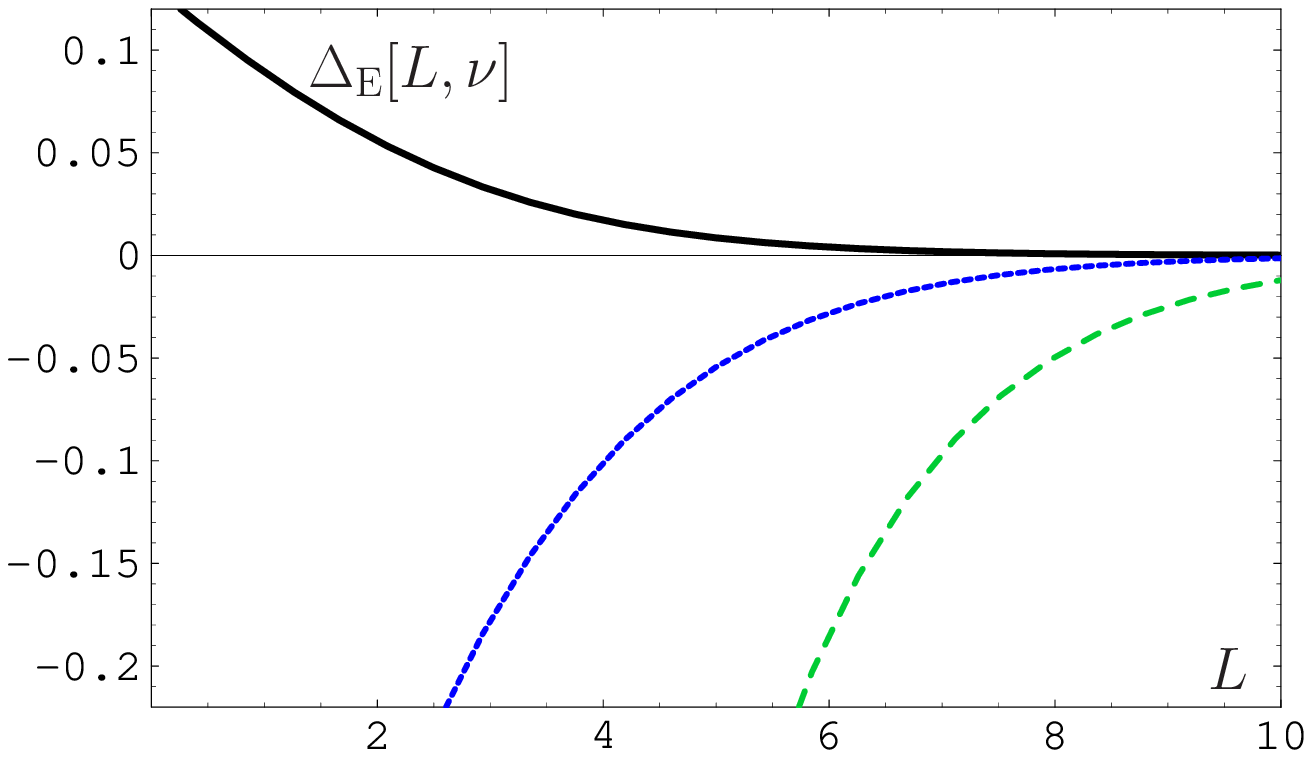}\vspace*{-1mm}}
   \caption{\small Left panel: Comparing \myMath{\mathfrak A_\nu}
   with \myMath{\left(\mathfrak A_{1}\right)^\nu}
   for fractional \myMath{\nu=0.62} (solid line), 1.62 (dotted line)
   and 2.62 (dashed line).
   Right panel: Comparing \myMath{\mathcal A_\nu}
   with \myMath{\left(\mathcal A_{1}\right)^\nu}
   for the same fractional values of \myMath{\nu} as in the left panel.
\label{fig:Delta.FAPT}}
\end{minipage}
\end{figure}

To demonstrate the importance of taking into account the FAPT,
that is using \myMath{\mathcal A_\nu[L]} and
\myMath{\mathfrak A_\nu[L]} instead of
\myMath{\left(\mathcal A_1[L]\right)^\nu} and
\myMath{\left(\mathfrak A_1[L]\right)^\nu},
we show in Fig.\ \ref{fig:Delta.FAPT}
the values of the normalized deviations
$\Delta_\text{M}(L,\nu)
 = 1- \left(\mathfrak A_{1}[L]\right)^{\nu}/\mathfrak A_{\nu}[L]$
and
$\Delta_\text{E}(L,\nu)
 = 1-\left(\mathcal A_{1}[L]\right)^{\nu}/\mathcal A_{\nu}[L]$
in the Minkowski and Euclidean domains,
respectively.

The construction of the FAPT with a fixed number of quark flavors, $N_f$,
is a two-step procedure:
we start with the perturbative result $\left[a(Q^2)\right]^{\nu}$,
generate the spectral density $\overline{\rho}_{\nu}(\sigma)$ using Eq.\ (\ref{eq:An.SD}),
and then obtain analytic couplings
$\overline{\mathcal A}_{\nu}[L]$ and $\overline{\mathfrak A}_{\nu}[L]$ via Eqs.\ (\ref{eq:A.U}).
Here $N_f$ is fixed and factorized out.
We can proceed in the same manner for $N_f$-dependent quantities:
$\left[\alpha_s^{}(Q^2;N_f)\right]^{\nu}$
$\Rightarrow$
$\rho_{\nu}(\sigma;N_f)=\rho_{\nu}[L_\sigma;N_f]
 \equiv\overline{\rho}_{\nu}(\sigma)/b_f^{\nu}$
$\Rightarrow$
$\mathcal A_{\nu}^{}[L;N_f]$ and $\mathfrak A_{\nu}^{}[L;N_f]$ ---
here $N_f$ is fixed but not factorized out.

The global version of the FAPT~\cite{BMS-FAPT}, (2009),
which takes into account heavy-quark thresholds,
is constructed along the same lines
but starting from global perturbative coupling
$\left[\alpha_s^{\,\text{\tiny glob}}(Q^2)\right]^{\nu}$,
being a continuous function of $Q^2$
due to choosing different values of QCD scales $\Lambda_f$,
corresponding to different values of $N_f$.
We illustrate here the case of only one heavy-quark threshold
at $Q^2=m_4^2$,
corresponding to the transition $N_f=3\to N_f=4$.
Then we obtain the discontinuous spectral density
\begin{eqnarray}
 \rho_n^\text{\tiny glob}(\sigma)
  = \theta\left(L_\sigma<L_{4}\right)\,
       \rho_n\left[L_\sigma;3\right]
  + \theta\left(L_{4}\leq L_\sigma\right)\,
       \rho_n\left[L_\sigma+\lambda_4;4\right]\,,~
 \label{eq:global_PT_Rho_4}
\end{eqnarray}
with $L_{\sigma}\equiv\ln\left(\sigma/\Lambda_3^2\right)$,
$L_{f}\equiv\ln\left(m_f^2/\Lambda_3^2\right)$
and
$\lambda_f\equiv\ln\left(\Lambda_3^2/\Lambda_f^2\right)$ for $f=4$,
which is expressed in terms of the fixed-flavor spectral densities
with 3 and 4 flavors,
$\rho_n[L_\sigma;3]$ and $\rho_n[L_\sigma+\lambda_4;4]$;
note here that
\myMath{L_\sigma+\lambda_4=\ln(\sigma/\Lambda_4^2)}.
However, it generates the continuous Minkowskian coupling
\begin{subequations}
\begin{eqnarray}
 {\mathfrak A}_{\nu}^{\text{\tiny glob}}[L]
  = \theta\left(L\!<\!L_4\right)
     \Bigl(\mathfrak A_{\nu}^{}[L;3]
          + \Delta_{43}\mathfrak A_{\nu}^{}
     \Bigr)
  + \theta\left(L_4\!\leq\!L\right)\,
     \mathfrak A_{\nu}^{}[L+\lambda_4;4]~~~~~
 \label{eq:An.U_nu.Glo.Expl}
\end{eqnarray}
with $\Delta_{43}\mathfrak A_{\nu}^{}=
            \mathfrak A_{\nu}^{}[L_4+\lambda_4;4]
          - \mathfrak A_{\nu}^{}[L_4;3]
$
and the analytic Euclidean coupling $\mathcal A_{\nu}^{\text{\tiny glob}}[L]$
\begin{eqnarray}
 \mathcal A_{\nu}^{\text{\tiny glob}}[L]
  = \mathcal A_{\nu}^{}[L+\lambda_4;4]
  + \int\limits_{-\infty}^{L_4}\!
       \frac{\overline{\rho}_{\nu}^{}[L_\sigma;3]
            -\overline{\rho}_{\nu}^{}[L_\sigma+\lambda_{4};4]}
            {1+e^{L-L_\sigma}}\,
         dL_\sigma\,.
  \label{eq:Delta_f.A_nu}
\end{eqnarray}
\end{subequations}

\begin{figure}[t!]
 \begin{minipage}{\textwidth}
  \centerline{\includegraphics[width=0.49\textwidth]{
  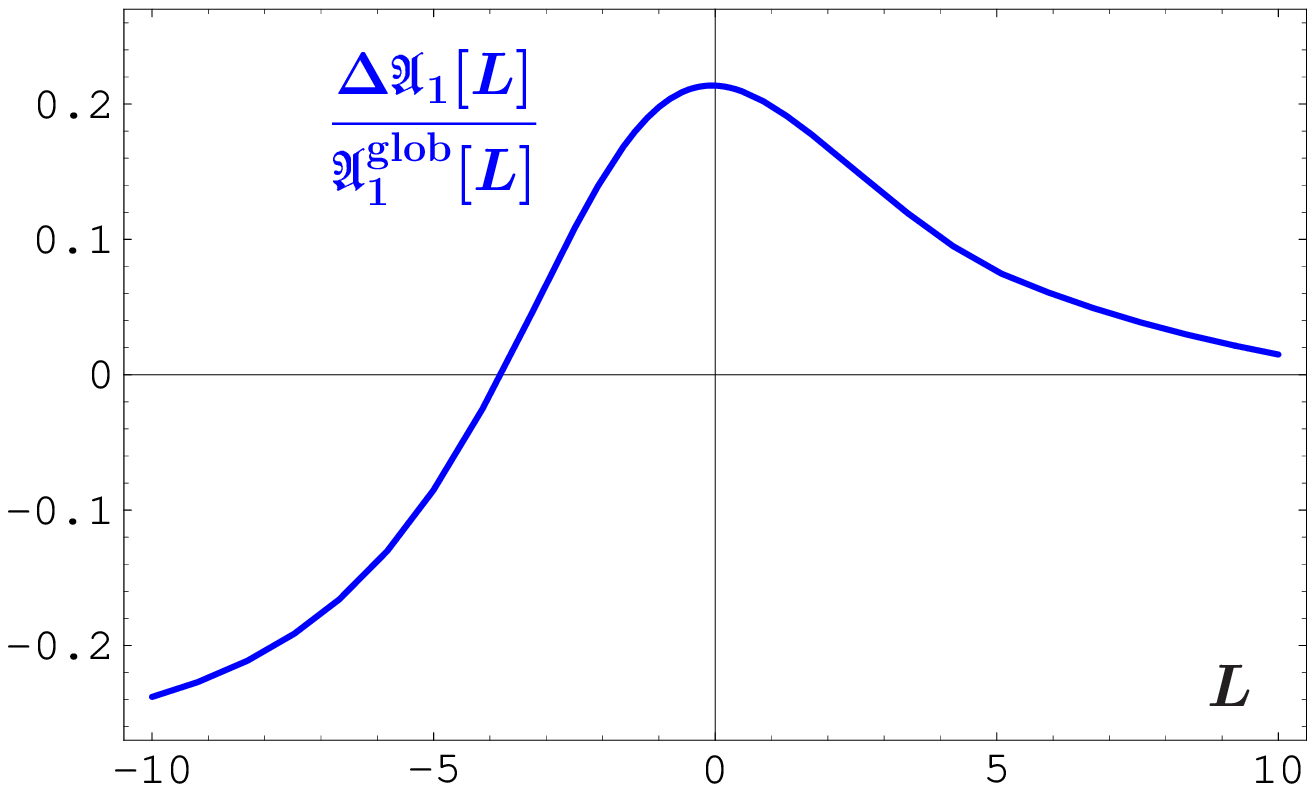}~\includegraphics[width=0.49\textwidth]{
  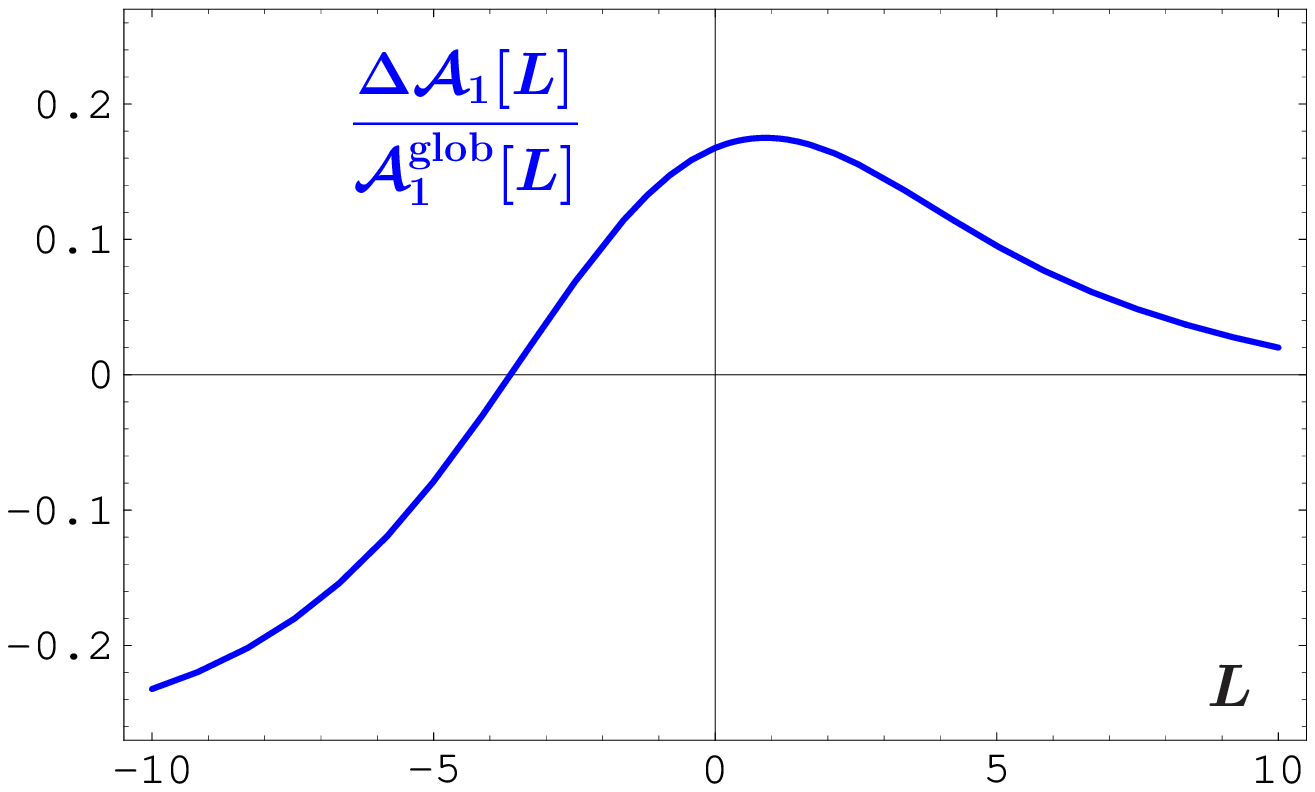}\vspace*{-1mm}}
   \caption{\small Left: Deviation of the global coupling relative
   to the fixed-$N_f$ coupling in the FAPT:
   $\Delta\mathfrak A_{1}[L]/\mathfrak A_{1}^\text{\tiny glob}[L]$.
   Right: The same for
   $\Delta\mathcal A_{1}[L]/\mathcal A_{1}^\text{\tiny glob}[L]$.
\label{fig:Delta.Global.A_1}}
\end{minipage}
\end{figure}
To demonstrate the magnitude of the threshold corrections,
we show in Fig.\ \ref{fig:Delta.Global.A_1}
the values of the normalized deviations
$\Delta\mathcal A_{\nu}[L]=
 \mathcal A_{\nu}^{\text{\tiny glob}}[L]
 -\mathcal A_{\nu}[L\!+\!\lambda_4;4]
$
and
$\Delta\mathfrak A_{\nu}[L]=
 \mathfrak A_{\nu}^{\text{\tiny glob}}[L]
 -\mathfrak A_{\nu}[L\!+\!\lambda_4;4]
$
in the Euclidean and Minkowski domains,
respectively
(for more details see~\cite{BMS-Resum}).

\subsection{Electromagnetic pion form factor at NLO}
 The scaled hard-scattering amplitude truncated at the next-to-leading order
(NLO)
and evaluated at renormalization scale $\mu_{R}^2=\lambda_{R} Q^2$
reads (see for details~\cite{BPSS04})
\begin{eqnarray}
 && T^\text{NLO}_\text{H}\left(x,y,Q^2;\mu_{F}^2,\lambda_\text{R}Q^2\right)
  = \alpha_s\left(\lambda_\text{R}Q^2\right)\,t_\text{H}^{(0)}(x,y)
 \nonumber\\
 \label{eq:T_Hard_NLO}
 && +\
    \frac{\alpha_s^2\left(\lambda_\text{R}Q^2\right)}{4\pi}
     \left\{C_\text{F}\,t_{\text{H},2}^{(1,\text{F})}\left(x,y;\frac{\mu_{F}^2}{Q^2}\right)
            + b_0\,t_\text{H}^{(1,\beta)}(x,y;\lambda_\text{R})
            + t_\text{H}^{(\text{FG})}(x,y)
     \right\}~~~
\end{eqnarray}
with shorthand notation ($\overline{x}\equiv1-x$)
\begin{eqnarray*}
 t_{\text{H},2}^{(1,\text{F})}\left(x,y;\frac{\mu_{F}^2}{Q^2}\right)
  = t_{\text{H}}^{(0)}\left(x,y\right)
     \left[2\Big(3 + \ln\,(\overline{x} \,\overline{y} \,) \Big)\ln\frac{Q^2}{\mu_{F}^2}\right]\,.~~~
\end{eqnarray*}
The leading twist-2 pion distribution amplitude (DA)~\cite{Rad77}
at the normalization scale $\mu_F^2$
is given by~\cite{ER80}
\begin{eqnarray*}
 \varphi_\pi(x,\mu_F^2)
  =  6\,x\,(1-x)
      \left[ 1
         + \sum\limits_{n\geq1}a_{2n}(\mu_F^2)\,C_{2n}^{3/2}(2 x -1)
      \right]\,.
\end{eqnarray*}
All {nonperturbative} information is encapsulated in the Gegenbauer coefficients
$a_{2n}(\mu^2_F)$.
To obtain the factorized part of the pion form factor (FF),
one needs to convolute
the pion DA with the hard-scattering amplitude:
\begin{eqnarray*}
 F_\pi^\text{Fact}(Q^2)
  = \varphi_\pi(x;\mu_{F}^2)\mathop{\otimes}\limits_{x}
      T^\text{NLO}_{H}\left(x,y;\mu_{F}^2,Q^2\right)
         \mathop{\otimes}\limits_{y}\varphi_\pi(y;\mu_{F}^2)\,.
\end{eqnarray*}
\begin{figure}[t!]
 \begin{minipage}{\textwidth}
  \centerline{\includegraphics[width=0.45\textwidth]{
  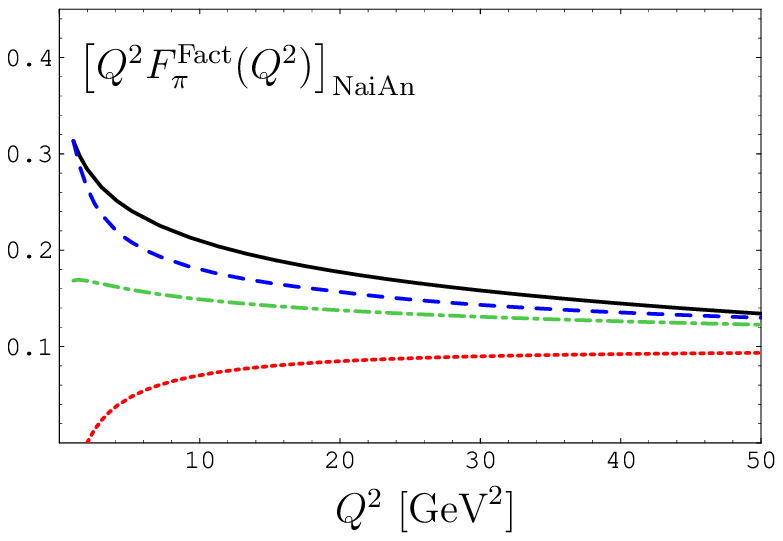}~~~\includegraphics[width=0.45\textwidth]{
  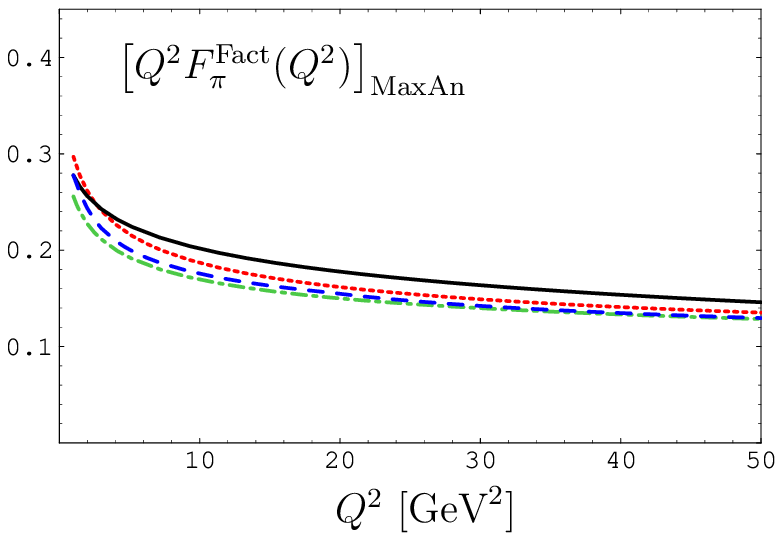}}
  \caption{\small Factorized pion FF in the ``Naive Analytization'' (left panel)
   and in the ``Maximal Analytization'' (right panel).
   Solid lines correspond to the scale setting $\mu_R^2=1$~GeV$^2$,
   dashed lines --- to $\mu_R^2=Q^2$,
   dotted lines --- to the BLM prescriptions,
   whereas dash-dotted lines --- to the $\alpha_v$-scheme. \label{fig:PionFF_Nai_Max}}
  \vspace*{-3mm}
\end{minipage}
\end{figure}
In order to obtain the analytic expression for the pion FF at the NLO
the so-called ``Naive Analytization'' was suggested in~\cite{SSK9900}.
It uses the analytic image ${\cal A}_{1}$ only for coupling itself
but not for its powers.
In contrast and in full accord with the APT ideology
the receipt of ``Maximal Analytization'' has recently been proposed
in~\cite{BPSS04}, which uses
the analytic image ${\cal A}_{2}$ for the second power of coupling
as well.
In Fig.\ \ref{fig:PionFF_Nai_Max} we show the predictions for
the factorized pion FF in the ``Naive'' and the ``Maximal Analytization''
approaches.
We see that in the ``Maximal Analytization'' approach
the obtained results are practically insensitive
to the renormalization scheme and scale-setting choice
(already at the NLO level).
It is interesting to note here that the FAPT approach,
used in~\cite{BMS-FAPT} for analytization of the $\ln(Q^2/\mu_{F}^2)$-terms
in the hard amplitude (\ref{eq:T_Hard_NLO}),
diminishes also the dependence on the factorization scale setting
in the interval $\mu_{F}^2=1-50$~GeV$^2$.

\begin{figure}[b!]
 \centerline{\includegraphics[width=0.5\textwidth]{
  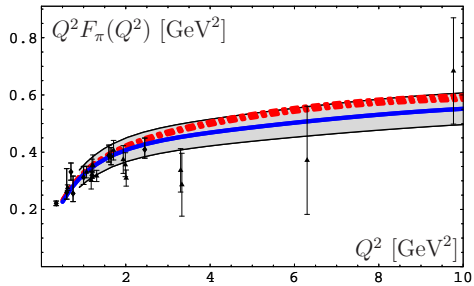}}
   \caption{\small We show as a narrow dashed-dotted strip the predictions
    for the pion FF obtained using the improved Gaussian model of the nonlocal QCD
    vacuum. The width of the strip is due to the variation of the Gegenbauer
    coefficients $a_2$ and $a_4$ in the corresponding shaded bands for the pion DA
    (indicated by the central solid line). Note that this dashed-dotted strip shows
    the effect of the $O(\mathcal A_2)$ correction only for the central solid curve
    of the shaded band. \protect\label{fig:FF.LD.2Loop}}
\end{figure}
Using the FAPT$\vphantom{^|}$ it appears to be possible to estimate
the NNLO correction to the whole pion FF
without very complicated calculation of the correspon\-ding
three-loop triangle spectral density~\cite{BPS09}.
Here we show in Fig.~\ref{fig:FF.LD.2Loop}
only the results:
The {NNLO} correction is of the order of \myMath{3-10\text{\%}}.

\subsection{Resummation in the one-loop APT and FAPT}
\label{sec:Resum.FAPT}
We consider now the perturbative expansion
of a typical physical quantity,
like the Adler function and the ratio $R$,
in the one-loop APT.
Due to limited space of our presentation
we provide all formulas only
for quantities in the Minkowski region:
\begin{eqnarray}
 \label{eq:APT.Series}
  \mathcal R[L]
   = \sum_{n=1}^{\infty}
      d_n\,\overline{\mathfrak A}_{n}[L]\,.
\end{eqnarray}
We suggest that there exists a generating function $P(t)$
for the coefficients $\tilde{d}_n=d_n/d_1$:
\begin{equation}
 \tilde{d}_n
  =\int_{0}^\infty\!\!P(t)\,t^{n-1}dt
   ~~~\text{with}~~~
   \int_{0}^\infty\!\!P(t)\,d t = 1\,.
 \label{eq:generator}
\end{equation}
To shorten our formulae, we use for the integral
$\int_{0}^{\infty}\!\!f(t)P(t)dt$
the following notation:
$\langle\langle{f(t)}\rangle\rangle_{P(t)}$.
Then the coefficients $d_n = d_1\,\langle\langle{t^{n-1}}\rangle\rangle_{P(t)}$
and, as has been shown in~\cite{MS04},
we have the exact result for the sum in (\ref{eq:APT.Series})
\begin{eqnarray}
 \label{eq:APT.Sum.DR[L]}
  \mathcal R[L]
   = d_1\,\langle\langle{\overline{\mathfrak A}_1[L-t]}\rangle\rangle_{P(t)}\,.
\end{eqnarray}
The integral in variable $t$ here has a rigorous meaning
ensured by the finiteness of the coupling
$\overline{\mathfrak A}_1[t]\leq 1$
and fast fall-off of the generating function $P(t)$.

In our previous publications~\cite{BMS-Resum},
we constructed generalizations of these results,
first, to the case of the global APT
when heavy-quark thresholds are taken into account.
Then one starts with the series
of type (\ref{eq:APT.Series}),
where $\overline{\mathfrak A}_{n}[L]$
are substituted by their global analogs
$\mathfrak A_{n}^\text{\tiny glob}[L]$
(note that due to different normalizations of global
 couplings,
$\mathfrak A_{n}^\text{\tiny glob}[L]\simeq\overline{\mathfrak A}_{n}[L]/\beta_f^n$,
 the coefficients $d_n$ should also be changed).
Then
\begin{eqnarray}
 \mathcal R^\text{\tiny glob}[L]
  &=&  d_1 \theta(L\!<\!L_4)
          \langle\langle{
           \Delta_{4}\mathfrak A_{1}[t]
           + \mathfrak A_{1}\!\Big[L\!-\!\frac{t}{\beta_3};3\Big]
           }\rangle\rangle_{P(t)}\nonumber\\
  &+& d_1  \theta(L\!\geq\!L_4)
          \langle\langle{
           \mathfrak A_{1}\!\Big[L\!+\!\lambda_4-\!\frac{t}{\beta_4};4\Big]
           }\rangle\rangle_{P(t)}\,;~~~
 \label{eq:sum.R.Glo.4}
\end{eqnarray}
where $\Delta_4\mathfrak A_\nu[t]\equiv
  \mathfrak A_\nu\!\Big[L_4+\lambda_{4}-t/\beta_4;4\Big]
 -\mathfrak A_\nu\!\Big[L_3-t/\beta_3;3\Big]$.

The second generalization has been obtained for the global FAPT.
Then the starting point is the series of the type
$\sum_{n=0}^{\infty} d_n\,\mathfrak A_{n+\nu}^\text{\tiny glob}[L]$
and the result of summation is a complete analog of Eq.\ (\ref{eq:sum.R.Glo.4})
with the substitutions
\begin{eqnarray}
 \label{eq:P_nu(t)}
  P(t)\Rightarrow P_{\nu}(t) =
   \int_0^{1}\!P\left(\frac{t}{1-x}\right)
    \frac{\nu\,x^{\nu-1}dx}
         {1-x}\,,
\end{eqnarray}
$d_0\Rightarrow d_0\,\mathfrak A_{\nu}[L]$,
$\mathfrak A_{1}[L-t]\Rightarrow \mathfrak A_{1+\nu}[L-t]$,
and
$\Delta_4\mathfrak A_{1}[t]\Rightarrow\Delta_4\mathfrak A_{1+\nu}[t]$.
All needed formulas have also been obtained
in parallel for the Euclidean case,
for details see~\cite{BMS-Resum}.

\subsection{Applications to Higgs boson decay}
\label{sec:Appl.Higgs}
Here we analyze the Higgs boson decay to a $\overline{b}b$ pair.
For its width we have
\begin{eqnarray}
 \label{eq:Higgs.decay.rate}
  \Gamma(\text{H} \to b\overline{b})
  = \frac{G_F}{4\sqrt{2}\pi}\,
     M_{H}\,
      \widetilde{R}_\text{\tiny S}(M_{H}^2)
\end{eqnarray}
with $\widetilde{R}_\text{\tiny S}(M_{H}^2)
     \equiv m^2_{b}(M_{H}^2)\,R_\text{\tiny S}(M_{H}^2)$
and
$R_\text{\tiny S}(s)$
is the $R$-ratio for the scalar correlator,
for details see~\cite{BMS-FAPT,BCK05}.
In the one-loop FAPT this generates the following
nonpower expansion\footnote{%
Appearance of denominators $\pi^n$ in association
with the coefficients $\tilde{d}_n$
is due to $d_n$ normalization.}:
\begin{eqnarray}
 \widetilde{\mathcal R}_\text{\tiny S}[L]
   =  3\,\hat{m}_{(1)}^2\,
      \Bigg\{\mathfrak A_{\nu_{0}}^{\text{\tiny glob}}[L]
          + d_1^\text{\,\tiny S}\,\sum_{n\geq1}
             \frac{\tilde{d}_{n}^\text{\,\tiny S}}{\pi^{n}}\,
              \mathfrak A_{n+\nu_{0}}^{\text{\tiny glob}}[L]
      \Bigg\}\,,
 \label{eq:R_S-MFAPT}
\end{eqnarray}
where $\hat{m}_{(1)}^2=9.05\pm0.09$~GeV$^2$ is the RG-invariant
of the one-loop $m^2_{b}(\mu^2)$ evolution
$m_{b}^2(Q^2) = \hat{m}_{(1)}^2\,\alpha_{s}^{\nu_{0}}(Q^2)$
with $\nu_{0}=2\gamma_0/b_0(5)=1.04$ and
$\gamma_0$ is the quark-mass anomalous dimension.
This value $\hat{m}_{(1)}^2$ was obtained
using the one-loop relation~\cite{KK08}
between the pole $b$-quark mass of~\cite{KuSt01}
and the mass $m_b(m_b)$.

We take for the generating function $P(t)$
the Lipatov-like model of~\cite{BMS-Resum}
with $\left\{c=2.4,~\beta=-0.52\right\}$
\begin{eqnarray}
 \label{eq:Higgs.Model}
  \tilde{d}_{n}^\text{\,\tiny S}
   = c^{n-1}\frac{\Gamma (n+1)+\beta\,\Gamma (n)}{1+\beta}
 \quad\text{with}\quad
  P_\text{\tiny S}(t)
   = \frac{(t/c)+\beta}{c\,(1+\beta)}\,e^{-{t/c}}\,.~~~
\end{eqnarray}
It gives a very good prediction for
$\tilde{d}_{n}^\text{\,\tiny S}$ with $n=2, 3, 4$,
calculated in the QCD PT~\cite{BCK05}:
$7.50$, $61.1$, and  $625$
in comparison with
$7.42$, $62.3$, and  $620$.
It is worthwhile to remind here the history
of calculating the \myMath{\beta}-function coefficients \myMath{\beta_n}
in the \myMath{\varphi^4_4} scalar field theory.
In~\cite{KST79} the resummation procedure was suggested
on the basis of taking into account four-loop results in the MOM scheme.
The five-loop calculations of the anomalous dimensions
\myMath{\gamma_2} and
\myMath{\gamma_4}
for this model in the \myMath{\overline{\text{MS}}} scheme
was performed in~\cite{ChKaTk81} (\myMath{\gamma_2})
and in~\cite{GLTC83}
(\myMath{\gamma_4} was calculated numerically
 with small errors).
In this last paper, using the Borel-like technique
and the four-loop results
in the \myMath{\overline{\text{MS}}} scheme
the five-loop prediction \myMath{\beta_5^\text{resum}=1405\pm 80}
was made.
The calculated result
\myMath{\beta_5=1420.69} appeared in the range
predicted in~\cite{KST79}.
The uncertainties of numerical calculations
were eliminated by Kazakov~\cite{Kaz84}
using the uniqueness method for  multiloop calculations
--- he confirmed numerical results.
After that, in~\cite{KNSFCL91},
the errors in the previous results for both \myMath{\gamma_2}
and \myMath{\gamma_4} were revealed.
As a result, \myMath{\gamma_2} and \myMath{\gamma_4} were changed,
but the value of \myMath{\beta_5} numerically
appears to be численно practically the same!
Resume: resummation predictions for the \myMath{\overline{\text{MS}}}-scheme
\myMath{\beta_5} are really in very good accord with the five-loop results.

Then we apply the FAPT resummation technique
to estimate
how good is the FAPT
in approximating the whole sum $\widetilde{\mathcal R}_\text{\tiny S}[L]$
in the range $L\in[11.5,13.7]$
which corresponds to the range
$M_H\in[60,180]$~GeV$^2$
with $\Lambda^{N_f=3}_{\text{QCD}}=189$~MeV
and ${\mathfrak A}^{\text{\tiny glob}}_{1}(m_Z^2)=0.122$.
In this range, we have ($L_6=\ln(m_t^2/\Lambda_3^2)$)
\begin{eqnarray}
 \frac{\widetilde{\mathcal R}_\text{\tiny S}[L]}
      {3\,\hat{m}_{(1)}^2}
  = {\mathfrak A}^\text{\tiny glob}_{\nu_{0}}[L]
   + \frac{d_{1}^\text{\,\tiny S}}{\pi}\,
      \langle\langle{\mathfrak A_{1+\nu_{0}}\!
                      \Big[L\!+\!\lambda_5\!-\!\frac{t}{\pi\beta_5};5\Big]
                     \Delta_{6}\overline{\mathfrak A}_{1+\nu_{0}}
                      \left[\frac{t}{\pi}\right]
      }\rangle\rangle_{P_{\nu_{0}}^\text{\,\tiny S}}~~~~~
 \label{eq:R_S.Sum}
\end{eqnarray}
with $P_{\nu_{0}}^\text{\,\tiny S}(t)$ defined via Eqs.\ (\ref{eq:Higgs.Model})
and (\ref{eq:P_nu(t)}).

\begin{figure}[h!]
 \centerline{\includegraphics[width=0.485\textwidth]{
 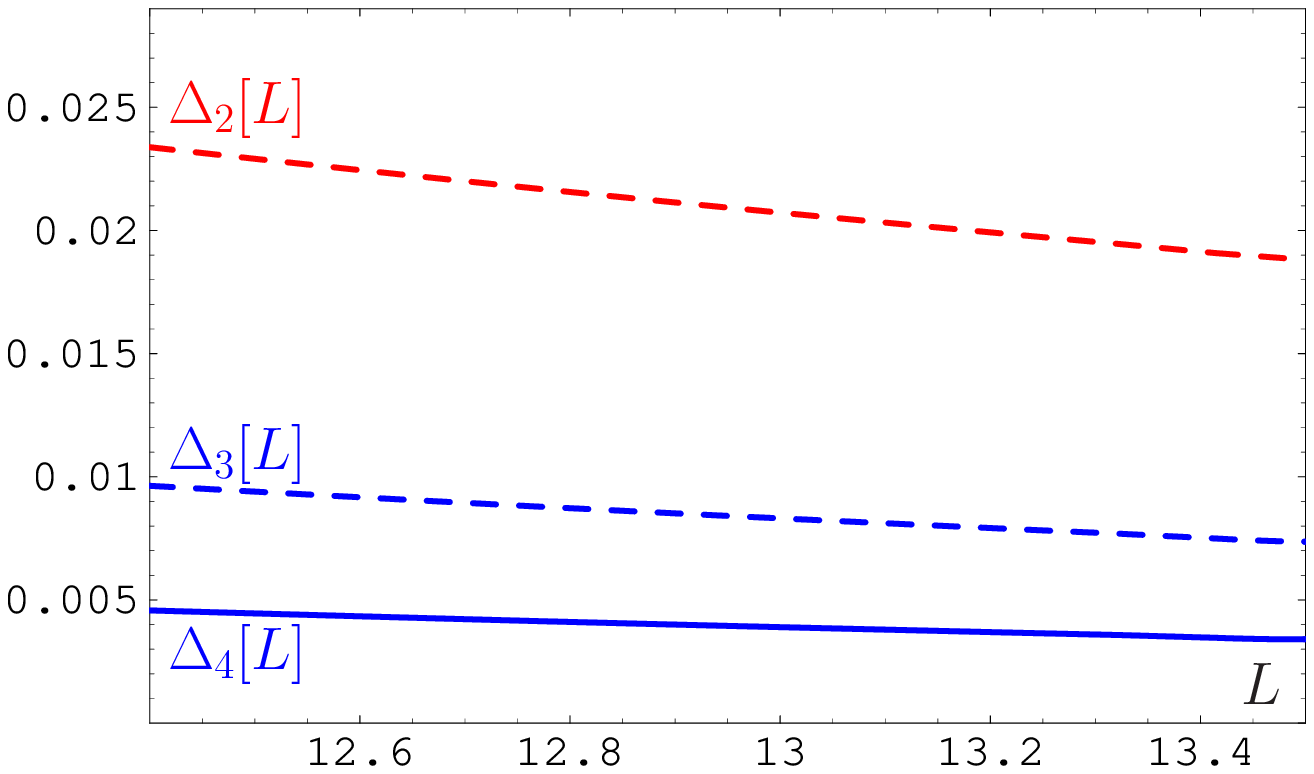}~~~\includegraphics[width=0.475\textwidth]{
 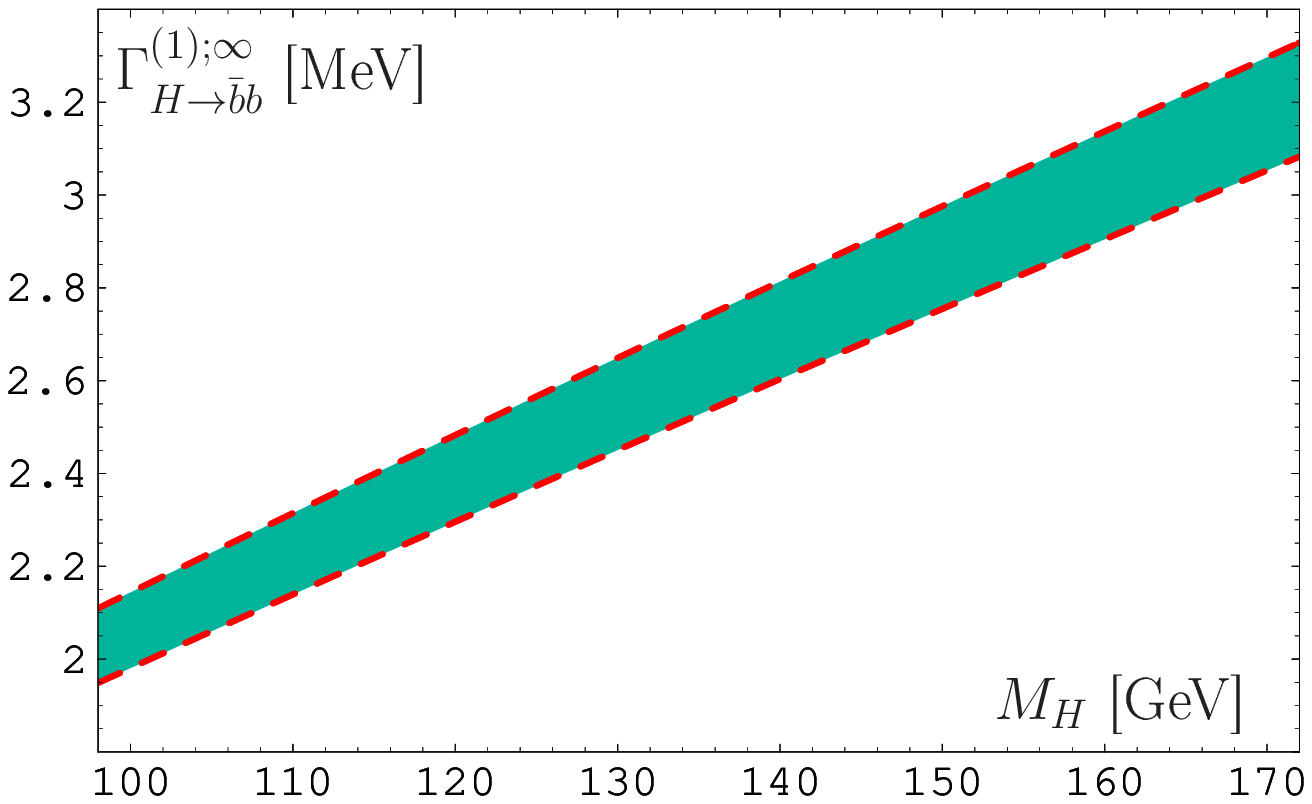}}
  \caption{\small Left panel: The relative errors $\Delta_N[L]$, $N=2, 3$
   and $4$, of the truncated FAPT in comparison with
   the exact summation result, Eq.\ (\ref{eq:R_S.Sum}).
   Right panel: The width $\Gamma_{H\to b\overline{b}}$ as a function
   of the  Higgs boson mass $M_{H}$ in the resummed FAPT.
   The width of the shaded strip is due to the overall uncertainties
   induced by the uncertainties of the resummation procedure
   and the pole mass error-bars.
   Both panels show the results obtained in the one-loop FAPT~\cite{BMS-Resum}.
   \label{fig:resum.1L}}
\end{figure}

Now we analyze the accuracy of the truncated FAPT expressions
\begin{eqnarray}
 \label{eq:FAPT.trunc}
 \widetilde{\mathcal R}_\text{\tiny S}[L;N]
  &=& 3\,\hat{m}_{(1)}^2\,
       \left[{\mathfrak A}_{\nu_{0}}^{\text{\tiny glob}}[L]
           + d_1^\text{\,\tiny S}\,\sum_{n=1}^{N}
              \frac{\tilde{d}_{n}^\text{\,\tiny S}}{\pi^{n}}\,
               {\mathfrak A}_{n+\nu_{0}}^{\text{\tiny glob}}[L]
       \right]
\end{eqnarray}
and compare them with the total sum
$\widetilde{\mathcal R}_\text{\tiny S}[L]$
in Eq.\ (\ref{eq:R_S.Sum})
using relative errors
$\Delta_N[L]=1-\widetilde{\mathcal R}_\text{\tiny S}[L;N]/\widetilde{\mathcal R}_\text{\tiny S}[L]$.
In Fig.~\ref{fig:resum.1L},
we show these errors for $N=2$, $N=3$, and $N=4$
in the analyzed range of $L\in[11,13.8]$.
We see that already $\widetilde{\mathcal R}_\text{\tiny S}[L;2]$
gives accuracy of the order of 2.5\%,
whereas $\widetilde{\mathcal R}_\text{\tiny S}[L;3]$
of the order of 1\%.
Looking at Fig.\ \ref{fig:resum.1L} we understand
that only in order to have the accuracy
better than 0.5\%,
one needs to take into account the 4-th correction.
We verified also that the uncertainty due to $P(t)$-modelling is small $\lesssim0.6\%$,
while the on-shell mass uncertainty is of the order of $2\%$.
The overall uncertainty then is of the order of $3\%$,
see in the right panel of Fig.\ \ref{fig:resum.1L},
that is in agreement with the Kataev\&Kim estimations~\cite{KK08}.

\section{Conclusions}
 \label{sec:Concl}
\begin{itemize}
 \item Perturbation approach in Quantum theory (both the
 nonrelativistic and QFT) suffers from divergence of
 (asymptotic) series in powers of small expansion
 parameter $g$. The reason is the nonanalyticity
 (essential singularity) at the origin of the $g$
 complex plane.\\
 Due to this, common power expansions have a lower
 limit of accuracy (as shown in Table 1).
 In particular, this refers to perturbative QCD. There,
 for some low-energy processes (see Table 2) this lower
 limit exceeds the experimental error. This is the
 reason that the nonperturbative means are of utmost
 importance in many physical situations.

 \item In Section 1, two of the nonperturbative
 instruments, the \textit{Renormalization Group} and the
 \textit{Dispersion Relation} methods, are outlined;
 description of their application to the actual case
 of Quantum Chromodynamics then follows. Here, the
 particular theoretical construction, the Analytic
 Perturbation Theory (APT), was devised on the turn of
 the century.

   \item  Section 2 begins with the summary of the APT
 basic elements. We also remind that in the APT one has

\begin{itemize}
 \item Universal (loop \& scheme independent) IR limit;
 \item Practical renormalization-scheme independence;
 \item Non-power perturbation expansion over a set of
 particular functions
 $\mathcal A_n(Q^2)\,,$ $\,\mathfrak A_n(s)\,$ instead
 of the running coupling powers $\alpha_s^n\,;$
 \item Quick loop convergence that improves the
 situation with the role of higher-loop corrections.
 \end{itemize}

 \item As a result of quick loop convergence, we show
 (see Table 3) that 3- and 4-loop APT terms contribute
 to observables less than 5\%, i.e., below the current
 level of data errors. Hence, the two-loop, NNLO level
 is practically sufficient.

 \item Then we expose the Fractional APT (FAPT) that
 provides an effective tool to apply the APT approach
 for renormgroup-improved perturbative amplitudes. By
 the example of the pion electromagnetic form factor
 we show that the FAPT delivers minimal sensitivity to
 both renormalization and factorization scale setting.

 \item In both the APT and FAPT approaches we describe
 the resummation procedures that produce finite resummed
 answers for perturbative quantities if one knows the
 generating functions $P(t)$ for the PT coefficients.
 Using quite simple model generating function
 \myMath{P_\text{S}(t)} for Higgs boson decay
 \myMath{H\to\overline{b}b} we conclude that at {N$^3$LO}
 we have accuracy of the order of 1\% due to the
 truncation error and of the order of 2\% due to the
 RG-invariant mass uncertainty.
\end{itemize}

\section*{Acknowledgements}
It is a pleasure to thank Drs. Sergey Mikhailov and Oleg Teryaev
for numerous discussions and useful remarks.
The fruitful remarks and discussions
with Drs. Konstantin Chetyrkin, Andrey Grozin,
Dmitry Kazakov, Andrey Kataev, Alexey Pivovarov, Anatoly Radyushkin,
and Nico Stefanis are also greatly acknowledged
by one of the authors (A.~B.).


\begin{thebibliography}{99}
\medskip

\begin{footnotesize} 

\bibitem{Dyson52}
 F.~J. Dyson,
  \textit{Phys. Rev.} \textbf{85}  (1952)  631.

\bibitem{KaSh80}
 D.~I. Kazakov and D.~V. Shirkov,
  \textit{Fortsch. Phys.} \textbf{28}  (1980)  465.

\bibitem{Lip76}
 L.~N. Lipatov,
  \textit{Sov. Phys. JETP} \textbf{45}  (1977)  216.

\bibitem{HuThPet52}
 Charles~Angas Hurst,
  \textit{Proc. Cambridge Phil. Soc.} \textbf{48}  (1952)  625;
 Walter~E. Thirring,
  \textit{Helv. Phys. Acta} \textbf{26}  (1953)  33;
 A.~Petermann,
  \textit{Phys. Rev.} \textbf{89}  (1953)  1160.

\bibitem{ShiDKS77}
 D.~V. Shirkov,
  \textit{Nuovo Cim. Lett.} \textbf{18}  (1977)  452;
 B.~D. Dorfel, D.~I. Kazakov, and D.~V. Shirkov,
  {JINR Preprint} E2-10720  (1977).

\bibitem{APTbasic}
 H.~F. Jones and I.~L. Solovtsov,
  \textit{Phys. Lett.} B349 (1995) 519 [hep-ph/9501344];
 D.~V. Shirkov and I.~L. Solovtsov,
  \textit{JINR Rapid Comm.} 2[76] (1996) 5 [hep-ph/9604363];
  \textit{Phys. Rev. Lett.} \textbf{79} (1997) 1209 [hep-ph/9704333];
 K.~A. Milton and I.~L. Solovtsov,
  Phys. Rev. D55  (1997)  5295 [hep-ph/9611438].

 \bibitem{RG-00}
  D.~V. Shirkov,
  ``Evolution of the Bogoliubov Renormalization Group'',
   in \textit{Quantum Field Theory, A XXth century profile},
   Ed. Asoke N.~Mitra,
   Hindustan Book Agency, INSA, New Dehli, 2000,  pp.~26--58
   [hep-th/9909024];
 V.~F. Kovalev and D.~V. Shirkov,
  \textit{Phys. Repts.} \textbf{352} (2001) 219--249
   [hep-th/0001210].

\bibitem{Gol07}
 G.~S.~Golitsyn,
  ``The portrait of an unknown person'',
   \textit{Priroda} \textbf{6} (2007) 61 (in Russian).

\bibitem{BS56}
 N.~N. Bogoliubov and D.~V. Shirkov,
  \textit{Nuovo Cim.} \textbf{3}  (1956)  845.

\bibitem{BSvvtkp}
 N.~N. Bogoliubov and D.~V. Shirkov,
  \textit{Introduction to the Theory of Quantum Fields}
  Wiley, New York, 1980, 720~p.

\bibitem{Dirac33}
 P.~A.~M. Dirac,
   in {\em Septi\'eme Conseil du Physique Solvay, Bruxelles,
   October 22--29, 1933},
    Ed. J. Cockcroft \textit{et~al.}
    (Gauthier-Villars, Paris, 1934), p.\ 203.

\bibitem{Shirkov92}
    D.~V. Shirkov,
    \textit{Nucl.Phys.} \textbf{B371} (1992) 467--481.

\bibitem{Kall52}
 G. K\"allen,
  \textit{Helv. Phys. Acta} \textbf{25} (1952) 417.

\bibitem{BLS60}
 N.~N. Bogoliubov, A.~A. Logunov, and D.~V. Shirkov,
  \textit{Sov. Phys. JETP} \textbf{10}  (1960)  574.

\bibitem{Bethke07}
 S.~Bethke,
  \textit{Prog. Part. Nucl. Phys.} \textbf{58}  (2007)  351
   [HEP-EX/0606035].

\bibitem{Khan11}
 R.~S. Pasechnik \textit{et al.},
  \textit{Phys. Rev.} \textbf{D78} (2008) 071902
  [ArXiv:0808.0066 [hep-ph]];
  \textit{Phys. Rev.} \textbf{D81} (2010) 016010
  [ArXiv:0911.3297 [hep-ph]];
 V.~L. Khandramai \textit{et al.}, paper in preparation.

\bibitem{SS06}
 D.~V. Shirkov and I.~L. Solovtsov,
  \textit{Theor. Math. Phys.} \textbf{150} (2007) 132
  [hep-ph/0611229].

\bibitem{Shi02lat}
 D.~V. Shirkov,
 \textit{Theor. Math. Phys.} \textbf{132} (2002) 1307--1317
  [hep-ph/0208082];
 G.~M. Prosperi, M. Raciti, and C. Simolo,
  Prog. Part. Nucl. Phys. 58  (2007)  387
   [hep-ph/0607209].

\bibitem{RKP82}
 A.~V. Radyushkin,
  \textit{JINR Rapid Commun.} \textbf{78}  (1996)  96
   [JINR Preprint, E2-82-159, hep-ph/9907228];
 N.~V. Krasnikov and A.~A. Pivovarov,
  \textit{Phys. Lett.} \textbf{B116}  (1982)  168.

\bibitem{MSSDVBRS}
 K.~A. Milton, I.~L. Solovtsov, and O.~P. Solovtsova,
  \textit{Phys. Lett.} \textbf{B415}  (1997)  104
  [hep-ph/9706409];
 Kimball~A. Milton and Olga~P. Solovtsova,
  \textit{Phys. Rev.} \textbf{D57}  (1998)  5402
  [hep-ph/9710316];
 I.~L. Solovtsov and D.~V. Shirkov,
  \textit{Phys. Lett.} \textbf{B442}  (1998)  344
  [hep-ph/9711251];
 A.~P. Bakulev, A.~V. Radyushkin, and N.~G. Stefanis,
  \textit{Phys. Rev.} \textbf{D62}  (2000)  113001
  [hep-ph 0005085];
 D.~V. Shirkov,
  \textit{Theor. Math. Phys.}\textbf{127}  (2001)  409
  [hep-ph/0012283];
  \textit{Eur. Phys. J.} \textbf{C22}  (2001)  331
  [hep-ph/0107282].

\bibitem{DVShi99}
 D.~V. Shirkov,
  \textit{Lett. Math. Phys.} \textbf{48}  (1999)  135.

\bibitem{Sim01}
 Y.~A. Simonov,
  \textit{Phys. Atom. Nucl.} \textbf{65} (2002) 135
  [hep-ph/0109081];
   ArXiv:1011.5386 [hep-ph].

\bibitem{BMS-FAPT}
 A.~P. Bakulev, S.~V. Mikhailov, and N.~G. Stefanis,
  \textit{Phys. Rev.} \textbf{D72}  (2005)  074014, 119908(E)
  [hep-ph/0506311];
  \textit{Phys. Rev.} \textbf{D75}  (2007)  056005;
                      \textbf{D77} (2008) 079901(E)
  [hep-ph/0607040];
 A.~P. Bakulev, A.~I. Karanikas, and N.~G. Stefanis,
  \textit{Phys. Rev.} \textbf{D72}  (2005)  074015
  [hep-ph/0504275];
 A.~P. Bakulev,
  \textit{Phys. Part. Nucl.} \textbf{40}  (2009)  715
  [arXiv:0805.0829 [hep-ph]];
 N.~G. Stefanis,
  ArXiv:0902.4805 [hep-ph].

\bibitem{MS04}
 S.~V. Mikhailov,
  \textit{JHEP} \textbf{0706}  (2007)  009
  [hep-ph/0411397].

\bibitem{BMS-Resum}
 A.~P. Bakulev and S.~V. Mikhailov,
  in {\em Proceedings of International Seminar on Contemporary Problems
      of Elementary Particle Physics,
      Dedicated to the Memory of I.~L.~Solovtsov,
      Dubna, January 17--18, 2008.},
  Eds. A.~P. Bakulev \textit{ et~al.}
  (JINR, Dubna, 2008), pp.\ 119--133
  [ArXiv:0803.3013 [hep-ph]];
 A.~P. Bakulev, S.~V. Mikhailov, and N.~G. Stefanis,
  \textit{JHEP} \textbf{1006}  (2010)  085
  [ArXiv:1004.4125 [hep-ph]].

\bibitem{BNPSS07}
 M. Baldicchi \textit{ et~al.},
  \textit{Phys. Rev. Lett.} \textbf{99}  (2007)  242001
  [ArXiv:0705.0329 [hep-ph]];
  \textit{Phys. Rev.} \textbf{D77}  (2008)  034013
  [ArXiv:0705.1695 [hep-ph]].

\bibitem{Mag10}
 B.~A. Magradze,
  \textit{Few Body Syst.} \textbf{48}  (2010)  143
  [ArXiv:1005.2674 [hep-ph]].

\bibitem{KS01}
 A.~I. Karanikas and N.~G. Stefanis,
  \textit{Phys. Lett.} \textbf{B504}  (2001)  225;
  \textit{ibid.} \textbf{B636} (2006) 330--331(E)
  [hep-ph/0101031].

\bibitem{BPSS04}
 A.~P. Bakulev  \textit{ et~al.},
  \textit{Phys. Rev.} \textbf{D70}  (2004)  033014
  [hep-ph/0405062].

\bibitem{Rad77}
 A.~V. Radyushkin,
  \uppercase{D}ubna preprint P2-10717, 1977
  [hep-ph/0410276].

\bibitem{ER80}
 A.~V. Efremov and A.~V. Radyushkin,
  \textit{Phys. Lett.} \textbf{B94}  (1980)  245.

\bibitem{SSK9900}
 N.~G. Stefanis, W. Schroers, and H.-C. Kim,
  \textit{Phys. Lett.} \textbf{B449}  (1999)  299;
  \textit{Eur. Phys. J.} \textbf{C18}  (2000)  137
  [hep-ph/0005218].

\bibitem{BPS09}
 A.~P. Bakulev, A.~V. Pimikov, and N.~G. Stefanis,
  \textit{Phys. Rev.} \textbf{D79}  (2009)  093010
  [ArXiv:0904.2304 [hep-ph]].

\bibitem{BCK05}
 P.~A. Baikov, K.~G. Chetyrkin, and J.~H. K{\"u}hn,
  \textit{Phys. Rev. Lett.} \textbf{96}  (2006)  012003
  [hep-ph/0511063].

\bibitem{KK08}
 A.~L. Kataev and V.~T. Kim,
  in {\em Proceedings of International Seminar on Contemporary Problems
  of Elementary Particle Physics, Dedicated to the Memory of I.~L.~Solovtsov,
  Dubna, January 17--18, 2008.},
  Eds. A.~P. Bakulev \textit{ et~al.}
  (JINR, Dubna, 2008), pp.\ 167--182
  [ArXiv:0804.3992 [hep-ph]];
  \textit{PoS} \textbf{ACAT08} (2009) 004.

\bibitem{KuSt01}
 Johann~H. K{\"u}hn and M. Steinhauser,
  \textit{Nucl. Phys.} \textbf{B619}  (2001)  588
  [hep-ph/0109084].

\bibitem{KST79}
 D.~I. Kazakov, D.~V. Shirkov, and O.~V. Tarasov,
  \textit{Theor. Math. Phys.} \textbf{38}  (1979)  9.

\bibitem{ChKaTk81}
 K.~G. Chetyrkin, A.~L. Kataev, and F.~V. Tkachov,
  \textit{Phys. Lett.} \textbf{B99}  (1981)  147.

\bibitem{GLTC83}
 S.~G. Gorishnii  \textit{ et~al.},
  \textit{Phys. Lett.} \textbf{B132}  (1983)  351.

\bibitem{Kaz84}
 D.~I. Kazakov,
  \textit{Phys. Lett.} \textbf{B133}  (1983)  406.

\bibitem{KNSFCL91}
 H. Kleinert \textit{ et~al.},
  \textit{Phys. Lett.} \textbf{B272}  (1991)  39
  [hep-th/9503230].

\end{footnotesize} 
\end{thebibliography}

\newcommand{\noopsort}[1]{} \newcommand{\printfirst}[2]{#1}
  \newcommand{\singleletter}[1]{#1} \newcommand{\switchargs}[2]{#2#1}

\end{document}